\documentclass[12pt, letterpaper]{article}
\usepackage[utf8]{inputenc}
\usepackage[T1]{fontenc}
\usepackage{amsmath}
\usepackage{amsfonts}
\usepackage{biblatex}
\usepackage{mathrsfs}
\usepackage{mathtools}
\usepackage{tikz}
\usetikzlibrary{matrix}
\numberwithin{equation}{section}
\usepackage{multirow}
\usepackage{hyperref}

\usepackage{varioref}
\usepackage[noabbrev]{cleveref}
\usepackage{float}
\usepackage{lmodern}
\usepackage{palatino}
\usepackage{mathptmx}
\usepackage[scaled=.90]{helvet}
\usepackage[affil-it]{authblk}
\usepackage{xcolor}

\begin{document}
\title{Open Topological String Amplitudes and BPS Invariants on Complete Intersection Calabi-Yau Threefolds }
\date{}
\author[]{Xuan Li \footnote{E-mail:\href{mailto: lixuan191@mails.ucas.ac.cn}{ lixuan191@mails.ucas.ac.cn}}, Yuan-Chun Jing and Fu-Zhong Yang}
\affil[]{School of Physical Sciences,University of Chinese Academy of Sciences,\\No.19(A) Yuquan Road, Shijingshan District, Beijing, P.R.China 100049}
\maketitle
\textit{\large This work is dedicated to the memory of our dear supervisor Prof. Fu-Zhong Yang, who passed away while this paper was being prepared.}
\begin{abstract}
    Open topological string partition function on compact Calabi-Yau threefolds satisfies the extended holomorphic anomaly equation. By direct integration, we solve these equations and obtain partition functions for first several genus and boundaries on complete intersection Calabi-Yau threefolds. Complemented by the unoriented worldsheet contribution, the annulus functions encode the  genus one BPS invariants. 
\end{abstract}

\newpage
\tableofcontents
\newpage

\section{Introduction}  
The study of topological strings on Calabi-Yau threefolds has led to important insights
in various aspects of string theory and supersymmetric gauge theories.  It has also benefited the development of enumerative problem on Calabi-Yau threefold by mirror symmetry\cite{Candelas1990,Kachru2000,Kachru1999,Katz2001,Li2001}. 

For the closed string theory, the string amplitudes are captured by the holomorphic anomaly equation \cite{Bershadsky1993a,Bershadsky1993}, which makes it possible to recursively determine the partition function at each genus up to a holomorphic ambiguity. Physically, the holomorphic anomaly equation can be interpreted as the realization of quantum background independence of the topological string, known as the wavefunction interpretation in \cite{Witten1993}. The non-holomorphic part of the topological string partition function  is generated by  finitely many generators\cite{Yamaguchi2004}, which is applied to solve high genus Gromov-Witten invariants on  one-modulus calabi-Yau threefolds\cite{Hosono2007,Huang2006}.
 
The  topological string partition functions in the presence of boundaries are first explored  in local Calabi-Yau.
The spacetime description of open string in terms of  string field  theory can be reduced to a matrix model\cite{Dijkgraaf2002b,Dijkgraaf2002,Dijkgraaf2002a}, which is generalized into the toric Calabi-Yau case in \cite{Bouchard2008,Bouchard2010,Marino2004,Marino2008}. It has developed into a mathematical technique appeaing in asymptotic expansion of many integrable system and enumertive problem, called topological recursion\cite{Eynard2014,Eynard2007a,Eynard2012,Eynard2007}. 

The open mirror symmetry relates topological A-model on  a Calabi-Yau threefold with A-branes as special Lagrangian submanifolds to topological B-model on a Calabi-Yau threefold with B-branes as holomorphic submanifolds. Homological mirror symmetry implies  the equivalence between the A-branes category, Fukaya category, and the  B-branes category, derived category of coherent sheaves\cite{Kontsevich1994,Kontsevich2000}. 
A-branes defined as the fixed locus of anti-holomorphic involution admit  two vacua  separated by a domainwall and the instantons are maps  from the disk to the Lagrangian A-branes.
The tree level data of open topological string theory is described by the domainwall tension, or the superpotential change between two domainwalls\cite{Witten1997}. It is the  solution of the inhomogeneous Picard-Fuchs differential equation\cite{Walcher2006}. as well as a Poincare normal function\cite{Morrison2007}, and encodes genus zero real BPS invariants\cite{Knapp2008,Krefl2008,Pandharipande2008,Walcher2009}.
The extended holomorphic anomaly equation for compact geometry with D-branes was proposed in \cite{Walcher2007},
which can be solved by direct integration using special geometry relation\cite{Walcher2007}, Feymann rule method\cite{Cook2007}, and polynomial structure of the partition function \cite{Alim2007,Konishi2007}. The high genus partition functions give rise to the prediction of high genus open Gromov-Witten invariants and real BPS invariants\cite{Knapp2008,Krefl2009,Shimizu2011,Walcher2007a}.

 In this article, we solve the high genus open topological partition function by directly integrating the extended holomorphic anomay equation on  one-modulus complete intersection Calabi-Yau (CICY) threefolds. In the large volume point, we extract the genus one real BPS invariants and list the several amplitudes for low boundaries and genus. The organization of this article are as follows:  In Second 2, we review the background knowledge about special geometry, partition functions, and holomorphic anomaly equation for closed and open topological string theory,  including the geometry of moduli space of the B-model, domaninwall tension, and the partition functions at tree, one-loop, and two-loop. In Section 3, the extended holomorphic anomaly equations are solved on four one-modulus CICY threefolds($X_{4,4}$,$X_{6,6}$,$X_{3,4}$,$X_{4,6}$), and genus one BPS invariants are extracted by including Klein bottle contribution to cancel tadpoles.  The last section is a brief summary and further discussion. In Appendix A, the BPS invariants of genus one are summarized. In Appendix B and C, the amplitudes determined by the domainwall tensions that are different from the domainwall tension given in section 3.1 and 3.2 are  calculated .

 \section{Special Geometry, Partition Functions ,and Holomorphic Anomaly Equation }
 \subsection{Special Geometry}
 The moduli space of the B-model is the complex structure moduli space $M_{CS}(X)$ of $X$. T he holomorphic $(3,0)$-form  $\Omega$  defines a line bundle $\mathcal{L}$ over $M_{CS}(X)$ with the metric,\begin{equation*}
     ||\Omega||^2=i \int \bar{\Omega}\wedge \Omega.
 \end{equation*}
The Zamolodchikov metric $G_{i\bar{j}}$  on $M_{CS}(X)$\cite{Zamolodchikov1986} , identified with the Weil-Peterson metric, is a Kahler metric given as the curvature of $\mathcal{L}$,
 \begin{equation}\label{eq:2.1}
   G_{i\bar{j}}=\partial_i\partial_{\bar{j}}K,
 \end{equation}
 with kahler potential 
 \begin{equation*}
 \quad  K=-\log  ||\Omega||^2.
 \end{equation*}
 In addition, there is a holomorphic symmetric tenor $C_{jkl}$ with coefficients in $\mathcal{L}^2$ satisfying,
 \begin{equation}\label{eq:2.2}
     \partial_{\bar{i}}C_{jkl}=0,\quad D_iC_{jkl}=D_j C_{kli},
 \end{equation}
 s.t., the Riemann curvature of $G_{i\bar{j}}$ reads,
 \begin{equation}\label{eq:2.3}
     R_{i\bar{j}k}^l\equiv -\partial_{\bar{j}}\Gamma^l_{ki}=G_{k\bar{j}}\delta_i^l+G_{i\bar{j}}\delta^l_k-e^{2K}C_{ikn}G^{m\bar{n}}C^*_{\bar{j}\bar{m}\bar{n}}G^{\bar{m}l}.
 \end{equation}
  Equation \ref{eq:2.1}, \ref{eq:2.2} and \ref{eq:2.3}
are known as  the $N=2$ special geometry relations on $M_{CS}$.
 
$G_{i\bar{j}}$ can also be written as,
\begin{equation*}
     G_{i\bar{j}}=\frac{g_{i\bar{j}}}{g_{0\bar{0}}},
 \end{equation*}
Here $g_{i\bar{j}}$ is the $tt^*$-metric on the vacuum bundle $\mathcal{V}\rightarrow X$,
\begin{equation*}
g_{i\bar{j}}=g(e_j,e_i)=\langle\Theta j| i\rangle,    
\end{equation*}
with  $\Theta$ is the CPT operator acting on the ground states. It induces  the $tt^*$-connection, i.e., the connection compatible with the metric and the holomorphic structure on $\mathcal{V}$, with connection matrix, $D_i(e_j)=(A_i)^k_j e_k$ and $A_i=g^{-1}\partial_i g$,
\begin{equation*}
 A_i=\left(
 \begin{tabular}{cccc}
 $g^{0\bar{0}}\partial_ig^{0\bar{0}}$&$0$&$0$&$0$\\
 $0$&$g^{\bar{j}l}\partial_ig_{m\bar{j}}$&$0$&$0$\\\
 $0$&$0$&$0$&$0$\\
 $0$&$0$&$0$&$0$\\
 \end{tabular}
 \right), \quad A_{\bar{i}}=\left(
 \begin{tabular}{cccc}
 $0$&$0$&$0$&$0$\\\
 $0$&$0$&$0$&$0$\\
 $0$&$0$&$g^{\bar{l}k}\partial_{bar{i}}g_{k\bar{m}}$&$0$\\
 $0$&$0$&$0$&$g^{0\bar{0}}\partial_{\bar{i}}g_{0\bar{0}}$\\
 \end{tabular}
 \right)
\end{equation*}
The $tt^*$-conncetion  and the chiral ring multiplication satisfy the $tt^*$-equations,
\begin{equation}
    \begin{gathered}
    [D_i,D_j]=[D_{\bar{i}},D_{\bar{j}}]=[D_i,C_{\bar{j}}]=[D_{\bar{i}},C_j]=0,\\
     [D_i,C_j]=[D_j,C_i],\quad [D_{\bar{i}},C_{\bar{j}}]=[D_{\bar{j}},C_{\bar{i}}],\\
     [D_i,D_{\bar{j}}]=-[C_i,C_{\bar{j}}],
    \end{gathered}
\end{equation}
 which  is equivalent to the flatness of the Gauss-Manin connection,
 \begin{equation*}
     \nabla_i=D_i-\alpha C_i,\quad \nabla_{\bar{i}}=D_{\bar{i}}-\alpha^{-1}C_{\bar{i}},
 \end{equation*}
 Here $C_i$ is the  action of the chiral fields over  $\mathcal{V}$, with matrix representation,
\begin{equation*}
 C_i=\left(
 \begin{tabular}{cccc}
 $0$&$0$&$0$&$0$\\
 $\delta^i_l$&$0$&$0$&$0$\\\
 $0$&$C_im^{\bar{l}}$&$0$&$0$\\
 $0$&$0$&$G_{i\bar{m}}$&$0$\\
 \end{tabular}
 \right),\quad C_{\bar{j}}=\left(
 \begin{tabular}{cccc}
 $0$&$G_{\bar{j}m}$&$0$&$0$\\\
 $0$&$0$&$C_{\bar{j}\bar{m}}^l$&$0$\\
 $0$&$C_im^{\bar{l}}$&$0$&$\delta^{\bar{l}}_{\bar{j}}$\\
 $0$&$0$&$0$&$0$\\
 \end{tabular}
 \right)
\end{equation*}

 The B-model on $X$ is determined by the variation of the complex structure of $X$,
 \begin{equation}\label{eq:2.5}
     H^{3-p,q}(X)\cong H^q(\wedge^pT X).
 \end{equation}
 The  cohomology groups $H^3(X)$ decomposes as follow,
 \begin{equation*}
  H^{3,0}(X)\oplus H^{2,1}(X)\oplus H^{1,2}(X)\oplus H^{0,3}(X)
 \end{equation*}
 which lead to a Hodge filtration on $H^3(X)$,
 \begin{equation*}
    H^{3,0}(X)=F^3H^{3}(X)\subset F^2H^3(X)\subset F^1H^3(X)\subset F^0H^3(X),
 \end{equation*}
 with
 \begin{equation*}
     F^q H^3(X)=\oplus_{q^\prime \geq q}H^{q^\prime,3-q^\prime}(X),
 \end{equation*}

On $M_{CS}(X)$, there are the  flat coordinates  defined by a symplectic basis of homology $3$-cycles, $\{\alpha_I, \beta^I \}_{I=0,1,\ldots,h^{1,2}}$.  The period
integrals of $\Omega$ are,
\begin{equation}
  \omega^I=\int_{\alpha_I}\Omega,\quad F_I=\int_{\beta^I}\Omega,
\end{equation}
Note that  $F_I$'s are homogeneous functions of $\omega^I$'s and can be expressed as derivatives of a single function $F(\omega)$, i.e.,
\begin{equation}
    F_I(\omega)=\frac{\partial F(\omega)}{\partial \omega^I}
\end{equation}
 where $F(\omega)$ is a homogeneous function of $\omega^I$'s of weight 2,called the prepotential. The flat coordinates of $M_{CS}$ is defined as the ratios of $\omega^I$'s,
 \begin{equation}
     t^i=\frac{\omega^i}{\omega^0},\quad i=1,\ldots,h^{1,2},
 \end{equation}
 which plays an important role in defining the mirror maps. Here  the generating function of the solutions of GKZ-system is used,
 \begin{equation*}
 \begin{aligned}
     &\varpi(z;\rho)=\sum_{n\geq 0} \frac{\Gamma[1-l_0(n+\rho)]}{\prod_{i>0}\Gamma[n+\rho+1]}z^{n+\rho},\\[10pt]
     &\omega^0=\varpi(z;0),\\[10pt]
     &\omega^1=\partial_{\rho}varpi(z;\rho)|_{\rho=0},\\[10pt]
     \ldots
     \end{aligned}
 \end{equation*}
 The prepotentials and the flat coordinates can be obtained  by solving the Picard-Fuchs equation governing the variation of Hodge structure. 
 
 The notion of special geometry has been generalized to the open string case, known as the N=1 special geometry\cite{Lerche2001,Lerche2002,Lerche2002a,Mayr2001}. In the open-closed B-model, the chiral ring is isomorphic to the relative cohomology group $H^3(X,D)$, with $D$ determined by the D-brane geometry. The bulk sector is identified with the section of $H^{3,3-p}(X)$ as Equation\ref{eq:2.5}, and the boundary sector
 is  given by the isomorphism,
 \begin{equation*}
     H^0(D,N_D)\cong H^{2,0}(D),\quad H^1(D,T|_D\wedge N_D)\cong H^{1,1}(D),
 \end{equation*}
 which can be interpreted as a Poincare residue for $\Omega$.
 The deformation of the open-closed B-model over the open-closed moduli space  is described by the variation of mixed Hodge structure on $H^3(X,D)$. The connection on the vacuum bundle is the Gauss-Manin connection on the relative cohomology bundle $\mathcal{H}^3(X,D)$, whose integrability and flatness determines the flat coordinates and superpotentials by Picard-Fuchs equations\cite{Forbes2003,Forbes2005}. 
 
 \subsection{Holomorphic Anomaly Equation}
The closed topological string partition funtion of genus one (torus) is found to be represented by a generalized index, \cite{Cecotti1993},
 \begin{equation*}
     \mathcal{F}^{(1)}=\frac{1}{2}\int \frac{d^2 \tau}{\tau_2}\mathrm{Tr}_{\mathrm{closed}}[(-1)^F F_L F_R e^{2\pi(\tau L_0- \bar{\tau}\bar{L}_0)}],
 \end{equation*}
 where the integral is defined over the fundamental domain of the action of $SL(2,\mathbb{Z})$ on the
upper half-plane. The torus one-point function can be obtained by the holomorphic differentiation of $\mathcal{F}^{(1)}$,
\begin{equation*}
    \partial_j \mathcal{F}^{(1)}=\frac{1}{2}\int \mathrm{Tr}_{\mathrm{closed}}(-1)^F [\int \mu G^-\int \bar{\mu}\bar{G}^- \phi_j(0)e^{2\pi i \tau(\tau L_0-\bar{\tau}\bar{L}_0)}],
\end{equation*}
which satisfy the following holomorphic anomaly equation at one-loop,
\begin{equation*}
    \partial_{\bar{i}}\partial_j\mathcal{F}^{(1)}=\frac{1}{2}TrC_{\bar{i}}C_{j}-\frac{1}{24}\mathrm{Tr}_{\mathrm{closeed}}(-1)^F G_{\bar{i}j}.
\end{equation*}

  The closed topological string patition function of genus  $g\geq 2$ is defined as the integral over the moduli space $M^{(g)}$ of Riemann surface $\Sigma^{(g)}$,
 \begin{equation*}
     \mathcal{F}^{(g)}=\int_{M^{(g)}}[dm]\langle\prod^{3g-3}_{a=1}(\int \mu_a G^-)(\int \mu_{\bar{a}}\bar{G}^-)\rangle_{\Sigma^{(g)}},\quad g\geq 2
 \end{equation*}
 where $\mu_a,a=1,\ldots,3g-3$ are the Beltrami diffrerentials.  It  satisfies the holomorphic anomaly equation \cite{Bershadsky1993} as follow,
 \begin{equation}
     \partial_{\bar{i}}\mathcal{F}^{(g)}=\frac{1}{2}\sum_{g_1+g_2=g}C^{jk}_{\bar{i}}\mathcal{F}^{(g_1)}_j\mathcal{F}^{(g_2)}_k +\frac{1}{2}C^{jk}_{\bar{i}}\mathcal{F}^{(g-1)}_{jk}, \quad g\geq 2,
 \end{equation}
 with $G^{jk}_{\bar{i}}\equiv C_{\bar{i}\bar{j}\bar{k}}g^{\bar{j}j}g^{\bar{k}k}=C_{\bar{i}\bar{j}\bar{k}}e^{2K}G^{\bar{j}j}G^{\bar{k}k}$.   On the right handside of the equation, the first term  originates from the closed string degeneration in which the Riemann
surface splits into two components, of genus $g_1(>0)$ and $g_2(>0)$, and  the second term comes
from the pinching of a handle that reduces the genus by one. 
 
 The amplitudes with insertion of chiral fields,
 \begin{equation*}
   \mathcal{F}^{(g)}_{i_1,\ldots,i_n}=\int_{M^{(g,0)}}[dm]\langle\int \phi^{(2)}_{i_1}\cdots \int \phi^{(2)}_{i_n} \prod^{3g-3}_{a=1}(\int \mu_a G^-)(\int \mu_{\bar{a}}\bar{G}^-)\rangle_{\Sigma_g},\quad g\geq 2  
 \end{equation*}
 can be obtained by covariant differentiation\cite{Dijkgraaf1990},
 \begin{equation*}
     \mathcal{F}^{(g)}_{i_1,\ldots,i_n,i_{n+1}}=D_{i_{n+1}}\mathcal{F}^{(g)}_{i_1,\ldots,i_n}
 \end{equation*}
 where $D$ is the Zamolodchikov-Kahler derivative on $\mathrm{Sym}^nT^*M\otimes\mathcal{L}^{2g-2}$. 
  
For open topological string, the partition function of one genus and zero boundaries satisfies the holomorphic anomaly equation that is slightly different to the closed string case,
\begin{equation}
    \partial_{\bar{i}}\mathcal{F}^{(1,0)}_j=\frac{1}{2}C_{jkl}C^{kl}_{\bar{i}}+(1-\frac{\chi}{24})G_{j\bar{i}},\\
\end{equation}
where the $1$ in the second term comes from the propagation of the unique ground state of zero charge $(q,\bar{q})=(0,0)$. The partition function of zero genus  and two boundaries (annulus or cylinder amplitude) is defined in \cite{Bershadsky1993},
 \begin{equation*}
     \mathcal{F}^{(0,2)}=\int^\infty_0\frac{dL}{L}\mathrm{Tr}[(-1)^F Fe^{-LH}],
 \end{equation*}
  which satisfies the holomorphic anomaly equation\cite{Walcher2007},
 \begin{equation}
     \partial_{\bar{i}}\mathcal{F}^{(0,2)}_j=-\Delta_{jk}\Delta^k_{\bar{i}}+\frac{N}{2}G_{j\bar{i}},
\end{equation}
where $N$ is the rank of a bundle over the moduli space in which the charge zero ground states of the open string live.
 
 The open topological string partition function $\mathcal{F}^{(g,h)}( 2g+h-2>0)$ is defined  as an integral over the moduli space $M^{(g,h)}$ of Riemann surface $\Sigma_{g,h}$of $g$ genus and $h$ boundaries,
 \begin{equation*}
     \mathcal{F}^{(g,h)}=\int_{M^{(g,h)}}[dm][dl]\langle \prod^{3g+h-3}_{a=1}(\int \mu_a G^-)(\int \bar{\mu}_{\bar{a}} \bar{G}^-)\prod^h_{b=1}\lambda_b(G^-+\bar{G}^-)\rangle_{\Sigma_{g,h}},
 \end{equation*}
 where $l^b$ is the length moduli and $m^a$ is the coordinates on $M^{(g,h)}$, and $\mu_a,\mu_{\bar{a}},a=1,\ldots,3g+3-h$ and $\lambda_b,b=1,\ldots,h$  are Beltrami differentials.
 The holomorphic anomaly equation in the presence of D-branes is as follow,
 \begin{equation} \label{eq:2.12}
     \partial_{\bar{i}}\mathcal{F}^{(g,h)}=\frac{1}{2}\sum_{\substack{g_1+g_2=g \\ h_1+h_2=h}}C^{jk}_{\bar{i}}\mathcal{F}^{g_1,h_1}_j\mathcal{F}^{g_2,h_2}_k+\frac{1}{2}C^{jk}_{\bar{i}}\mathcal{F}^{(g-1,h)}_{jk}-\Delta^j_{\bar{i}}\mathcal{F}_j^{(g,h-1)},
 \end{equation}
with $\Delta^j_{\bar{i}}\equiv \Delta_{\bar{i}\bar{j}}g^{\bar{j}j}=\Delta_{\bar{i}\bar{j}}e^{2K}G^{\bar{j}j}$. 
Again, the first two terms come from closed string degenerations, while the last comes
from the shrinking of a boundary component to zero size. Degenerations in the open
string channel do not contribute generically. 
The amplitudes with insertion is defined as,
 \begin{equation*}
     \mathcal{F}^{(g,h)}_{i_1,\ldots,i_n}=\int_{\mathcal{M}^{(g,h)}}\langle \int \phi^{(2)}_{i_1}\cdots\int \phi^{(2)}_{i_n} \prod^{3g+h-3}_{a=1}(\int \mu_a G^-)(\int \bar{\mu}_{\bar{a}} \bar{G}^-)\prod^h_{b=1}\lambda_b(G^-+\bar{G}^-)\rangle_{\Sigma_{g,h}},
 \end{equation*}
again  given by the covariant differentiaiton ,
 \begin{equation*}
     \mathcal{F}^{(g,h)}_{i_1,\ldots,i_n,i_{n+1}}=D_{i_{n+1}} \mathcal{F}^{(g,h)}_{i_1,\ldots,i_n}.
 \end{equation*}
 
 The equation \ref{eq:2.12}  is solved by Feymann rules. In this case, the propagators $S,S^i,S^{ij}$ relate to the Yukawa coupling,
\begin{equation}\label{eq:2.13}
    \partial_{\bar{i}}S^{ij}=C_{\bar{i}},\quad \partial_{\bar{i}}S^j=G_{i\bar{i}}S^{ij},\quad \partial_{\bar{i}}S=G_{i\bar{i}}S^i,
\end{equation}
which are sections of the bndles $\mathcal{L}^{-2}\otimes \mathrm{Sym}^m T$, and $\Delta,\Delta^i$ relate to the disk function,
\begin{equation}\label{eq:2.14}
    \partial_{\bar{i}}\Delta^j=\Delta_{\bar{i}}^j,\quad \partial_{\bar{i}}\Delta=G_{i\bar{i}}\Delta^i.
\end{equation}
which are sections of the bndles $\mathcal{L}^{-1}\otimes \mathrm{Sym}^m T$. The vertices of the Feymann rules are given by the correlation function $\mathcal{F}^{(g,h)}_{i_1,\ldots,i_n}$.

In this paper, we study one-parameter models in weighted projective space.  The propagators  in the holomorphic limits are as follow,
\begin{equation}\label{eq:2.15}
\begin{aligned}
  &  S^{zz}=\frac{1}{C_{zzz}}\partial_z \log(G^{z\bar{z}}(ze^K)^2),\\[10pt]
 &   S^z=\frac{1}{C_{zzz}}[(\partial_z\log(ze^K))^2-D_z\partial_z\log(ze^K)],\\[10pt]
  &  S=[S^z-\frac{1}{2}D_zS^{zz}-\frac{1}{2}(S^{zz})^2C_{zzz}]\partial_z\log(ze^K)+\frac{1}{2}D_zS^z+\frac{1}{2}S^{zz}S^zC_{zzz}\\[10pt]
  &  \Delta^z =-\Delta_{zz}C_{zzz}^{-1},\\[10pt]
  & \Delta=D_z\Delta^z.
\end{aligned}
\end{equation}
 \subsection{Open String at Tree Level}
 The BPS domainwall tension plays an important role in the open topological string theory  at tree level. and open mirror symmetry relates the domainwall tension in the A-model   to the domainwall tension in the  B-model,
 \begin{equation*}
     \mathcal{T}_A(t)=\varpi_0(z(t))^{-1}\mathcal{T}_B(z(t)).
 \end{equation*}
 
 In the A–model, $\mathcal{T}_A(t)$ is defined as the generating functional counting holomorphic disks ending on the A-brane $L_\alpha$, with the form
\begin{equation}
  \mathcal{T}_A(t)=\frac{t}{2}+\mathcal{T}_{\mathrm{classical}}+\sum_{D\in H_2(X,L_\alpha,\mathbb{Z})} \tilde{n}_D q^{\mathrm{Area(D)}},  
\end{equation}
where $H_2(X,L_\alpha, \mathbb{Z})$ is the relative cohomology group labeling the classes $D$ of the image of the holomorphic discs. 

In the B–model, $\mathcal{T}_B(z)$ is the superpotential change on two domainwalls formed by two D5-branes  wrapped on two distinct holomorphic curves $C_+$ and $C_-$ in the same homology class \cite{Witten1997}, 
\begin{equation*}
    \mathcal{T}=\int_\Gamma \Omega,
\end{equation*}
 where $\Gamma$ is a three chain with boundary $\partial \Gamma =C_+-C_-$. Mathematically speaking, it is known as a Abel-Jacobi map\cite{Aganagic2001,Aganagic2000,Li2009}, and
 a Poincare Poincare normal function that should be viewed as a holomorphic section of the Griffiths intermediate Jacobian fibration over $M_{CS}(X)$\cite{Green1994,voisin2002,voisin2003}.  This domainwall tension satisfies the inhomogeneous Picard-Fuchs equation,
 \begin{equation}
     \mathcal{L}_{PF}\mathcal{T}_B(z)=f(z),
 \end{equation}
 where $\mathcal{L}_{PF}$ is the Picard-Fuchs equation  from variation of Hodge structure \cite{Morrison1993,Schmid1973} by Griffiths-Dwork method\cite{Griffiths1969,Griffiths1969a} or GKZ-system\cite{Hosono1995,Hosono1993,Hosono1994,Gelfand1990}.

In the tree level,  the most fundamental amplitudes of open topological string are the Yukawa  $C_{ijk}$ and the disk amplitudes  $\Delta_{ij}$.
 The Yukawa couplings is defined as,
\begin{equation*}
    C_{ijk}=-\langle \Omega,\nabla_i \nabla_j \nabla_k \Omega\rangle=-\langle\Omega,\partial_i\partial_j\partial_k\Omega\rangle.
\end{equation*}
where $\Omega(z)\in H^{3,0}(X)$ is the holomorphic three form on the Calabi-Yau threefold $X$, $\nabla$  the Gauss-Manin connection satisfying the Griffiths transversality, and $\langle\cdot,\cdot\rangle=\int_{X} \cdot\wedge\cdot$ the symplectic pairing on $H^3(X)$. The Yukawa coupling is holomorphic and can be written as the covariant differentiation of the genus zero partition function $\mathcal{F}^{(0)}$,
\begin{equation*}
\partial_{\bar{i}}C_{ijk}=0,\quad C_{ijk}=D_iD_jD_k\mathcal{F}^{0}.
\end{equation*}
In particular, the Yuakawa couplings of  the one-modulus models in this article have the form \cite{Batyrev1995},
 \begin{equation}\label{eq:2.18}
     C_{zzz}=\frac{W(0)}{(1-\mu z)\varpi_0}(\frac{dz}{z})^{\otimes 3}.
 \end{equation}
 
Meanwhile, the disk amplitude is identified with the Griffiths infinitesimal invariant\cite{Green1989,Griffiths1983},
\begin{equation*}
   \mathcal{F}^{(0,1)}_{ij}= \Delta_{ij}=\langle \Omega, \nabla_i \nabla_j \nu\rangle
\end{equation*}
with $\nu$  a  normal function defined by a three-chain $\Gamma\subset X$.
By the Griffiths transversality condition of the normal function $\nu$, i.e., $\langle\Omega,\nabla \nu\rangle=0$, $\Delta_{ij}$ can be rewritten by covariant differentiation,
 \begin{equation*}
     \Delta_{ij}=D_i D_j\mathcal{T}-C_{ijk}g^{k\bar{k}}D_{\bar{k}}\bar{\mathcal{T}}.
 \end{equation*} 
 Instead of being holomorphic, $\Delta_{ij}$  satisfies the following equations,
\begin{equation}\label{eq:2.19}
    \partial_{\bar{k}}\Delta_{ij}=-C_{ijl}\Delta^l_{\bar{k}},\quad \Delta^k_{\bar{i}}=\Delta_{\bar{i}\bar{j}}e^KG^{k\bar{j}}.
\end{equation}
and in the holomorphic limit,
 \begin{equation}\label{eq:2.20}
     \lim_{\bar{z}\rightarrow 0}\Delta_{ij}=     \lim_{\bar{z}\rightarrow 0}D_iD_j\mathcal{T}=\partial_z\partial_z \mathcal{T}.
\end{equation}

\subsection{Open String at One-loop}
The torus amplitudes can  be solved by equation \ref{eq:2.13},
 \begin{equation*}
 \begin{aligned}
     \partial_{\bar{i}} \mathcal{F}^{(1,0)}&=\frac{1}{2}C_{jkl}C^{kl}_{\bar{i}}+(1-\frac{\chi}{24})G_{\bar{j}i}\\
     &=\frac{1}{2}C_{jkl}\partial_{\bar{i}}S^{kl}+(1-\frac{\chi}{24})\partial_j\partial_{\bar{i}}K\\
    & =\partial_{\bar{i}}(\frac{1}{2}C_{jkl}S^{kl}+(1-\frac{\chi}{24})K_i)\\
     \end{aligned}
 \end{equation*}
 i.e.,
 \begin{equation*}
  \mathcal{F}^{(1,0)}=\frac{1}{2}C_{jkl}S^{kl}+(1-\frac{\chi}{24})K_i+hol.amb,
 \end{equation*}
 which has the following general form after inserting the propagators $S^{kl}$,
 \begin{equation*}
     \mathcal{F}^{(1,0)}=\frac{1}{2}\log[\det G_{\bar{i}j}^{-1}e^{K(3+n-\frac{1}{12}\chi)}|hol.amb.|^2],
 \end{equation*}
 In the holomorphic limit, this amplitude at large volume on one -parameter is,
\begin{equation}\label{eq:2.21}
    \mathcal{F}^{(1,0)}\underset{hol.lim.}{\rightarrow}\frac{1}{2}\log[(\frac{q}{z}\frac{dz}{dq})\varpi_0^{\frac{\chi}{12}-4}z^{-\frac{c_2}{12}}(discirminant)^{-\frac{1}{6}}]
\end{equation}
with $\chi$ the Euler number and  $c_2$ the second chern class of the Calabi-Yau threefolds.

Similarly, the annulus amplitudes can  be solved by equation \ref{eq:2.14},
 \begin{equation*}
 \begin{aligned}
     \partial_{\bar{i}}\partial_j \mathcal{F}^{(2,0)}&=\partial_{\bar{i}}(-\Delta_{jk}\Delta^k+\frac{N}{2}\partial_j K)-C_{jkl}\Delta^l_{\bar{i}}\Delta^k\\
     &=\partial_{\bar{i}}(-\Delta_{jk}\Delta^k-\frac{1}{2}C_{jkl}\Delta^k\Delta^l+\frac{N}{2}\partial_i K )\\
     &=\partial_{\bar{i}}(-\frac{1}{2}(\Delta_{jk}+f_{jk})\Delta^k+\frac{N}{2}\partial_j K)
     \end{aligned}
 \end{equation*}
 i.e.,
 \begin{equation*}
  \mathcal{F}^{(2,0)}_j=-\frac{1}{2}\Delta_{jk}\Delta^k+\frac{N}{2}K_j+hol.amb.
 \end{equation*}
 or
  \begin{equation}\label{eq:2.22}
  \mathcal{F}^{(2,0)}_z=-\frac{1}{2}\Delta_{zz}\Delta^z+hol.amb.
 \end{equation}
 for one-parameter models.
Then, $\mathcal{F}^{(0,2)}(z)_B$ can be obtained in the B-model, and relates to $\mathcal{F}^{(0,2)}_A$ \footnote{Here A/B refer to A/B-model, not a insertion }by 
\begin{equation*}
     \mathcal{F}^{(0,2)}_A(t)=\mathcal{F}^{(0,2)}_B(z),
\end{equation*}
where $\mathcal{F}^{(0,2)}_A(t)$ is defined as the generating functional counting holomorphic maps of annuli ending on $L$, with the form,
 \begin{equation*}
     \mathcal{F}^{(0,2)}_A=\sum_{A\in H_2(X,L,\mathbb{Z})}\tilde{n}_A q^{\mathrm{Area(A)}},
 \end{equation*}
where $\tilde{n}_A$ are not integer in general. Beacuse the A- and B- model only decouple if the tadpoles are cancelled, the integrality of BPS state counting can be assured only when unoriented worldsheets are included. Specificially, the generating functional $\mathcal{K}$ for the  holomorphic maps of Klein bottles in the A-model is,
\begin{equation*}
    \mathcal{K}_A(t)=\sum_{K\in H_2(X,L,\mathbb{Z})}n_Kq^{\mathrm{Area(K)}}.
\end{equation*}
The corresponding B-model Klein bottle partition function satisfies the holomorphic anomaly equation,
\begin{equation*}
    \partial_{\bar{i}}\partial_j \mathcal{K}=\frac{1}{2}C^{kl}_{
\bar{i}}C_{jkl}-G_{\bar{i}j}
\end{equation*}
which can be solved by special geometry relation \ref{eq:2.3},
\begin{equation*}
    \mathcal{K}=\frac{1}{2}\log(\det G_{\bar{i}j}^{-1}e^{K(n-1)}|hol.amb.|^2).
\end{equation*}
It turns out that under holomorphic limit, above equation can be rewritten for one-paramter models as,
\begin{equation}\label{eq:2.23}
    \mathcal{K}\rightarrow \frac{1}{2}\log[(\frac{q}{z}\frac{dz}{dq})(discrimnant)^{-1/4}],
\end{equation}
with the expansion at the large volume point,
\begin{equation*}
    \mathcal{K}\underset{hol.lim.}{\rightarrow}\sum_{\text{d even}}\tilde{n}_d^{(1,0)_k}q^{d/2}.
\end{equation*}

The one-loop open/unoriented partition function receive non-trivial contribution from annulus partition functions $\mathcal{A}=\mathcal{F}^{(0,2)}(z)$ and Klein bottle partition functions $\mathcal{K}$, underlying the real BPS invariants $n_d^{(1,real)}$,
\begin{equation}\label{eq:2.24}
    \mathcal{A}+\mathcal{K}=2\sum_{\substack{\text{d even}\\ \text{k odd}}}\frac{1}{k}n_d^{(1,real)}q^{dk/2}.
\end{equation}

\subsection{Open String at Two-loop}
Similarly , for $(g,h)=(0,3)$, we use equation \ref{eq:2.19},
  \begin{equation*}
  \begin{aligned}
 \partial_{\bar{i} }\mathcal{F}^{(0,3)}&=-\Delta^j_{\bar{i}}\mathcal{F}^{(0,2)}_j\\
 &=-\partial_{\bar{i}}(\Delta^j\mathcal{F}^{(0,2)}_j)+(-\Delta_{jk}\Delta^k_{\bar{i}}+\frac{N}{2}G_{j\bar{i}})\Delta^j\\
 &=-\partial_{\bar{i}}(\Delta^j\mathcal{F}^{(0,2)}_j)-\Delta_{jk}\Delta^k_{\bar{i}}\Delta^j+\frac{N}{2}G_{j\bar{i}}\Delta^j\\
  &=-\partial_{\bar{i}}(\Delta^j\mathcal{F}^{(0,2)}_j)-\frac{1}{2}\partial_{\bar{i}}(\Delta_{jk}\Delta^k\Delta^j)-\partial_{\bar{i}}\Delta_{jk}\Delta^j\Delta^k+\frac{N}{2}G_{j\bar{i}}G^{j\bar{j}}\partial_{\bar{j}}\Delta\\
    &=-\partial_{\bar{i}}(\Delta^j\mathcal{F}^{(0,2)}_j)-\frac{1}{2}\partial_{\bar{i}}(\Delta_{jk}\Delta^k\Delta^j)-C_{jkl}\Delta^l_{\bar{i}}\Delta^j\Delta^k+\frac{N}{2}\partial_{\bar{i}}\Delta\\
     &=\partial_{\bar{i}}(-\Delta^j\mathcal{F}^{(0,2)}_j+\frac{N}{2}\Delta-\frac{1}{2}\Delta_{jk}\Delta^k\Delta^j-\frac{1}{6}C_{jkl}\Delta^l\Delta^j\Delta^k)\\
  \end{aligned}
 \end{equation*}
 i.e.,
 \begin{equation*}
     \mathcal{F}^{(0,3)}=-\mathcal{F}^{0,2}_j\Delta^j+\frac{N}{2}\Delta-\frac{1}{2}\Delta_{jk}\Delta^j\Delta^k-\frac{1}{6}C_{jkl}\Delta^j\Delta^k\Delta^l+hol.amb.
 \end{equation*}
 or
 \begin{equation}\label{eq:2.25}
     \mathcal{F}^{(0,3)}=-\mathcal{F}^{0,2}_z\Delta^z-\frac{1}{3}\Delta_{zz}\Delta^z\Delta^z+hol.amb.
 \end{equation}
 for one-parameter models

The oriented one-loop partition function is given by,
  \begin{equation*}
 \begin{aligned}
     \partial_{\bar{i}} \mathcal{F}^{(1,1)}&=\frac{1}{2}C^{jk}_{\bar{i}}\Delta_{jk}-\mathcal{F}^{(1,0)}\Delta^i_{\bar{i}}\\
     &=\partial_{\bar{i}}(\frac{1}{2}S^{jk}\Delta_{jk}-\mathcal{F}^{(1,0)}_j\Delta^j)+\frac{1}{2}S^{jk}C_{jkl}\Delta^l_{\bar{i}}+(\frac{1}{2}C_{jkl}C^{kl}_{\bar{i}}-(\frac{\chi}{24}-1)G_{j\bar{i}})\Delta^j\\
     &=\partial_{\bar{i}}(\frac{1}{2}S^{jk}\Delta_{jk}-\mathcal{F}^{(1,0)}_j\Delta^j+\frac{1}{2}C_{jkl}S^{kl}\Delta^l-(\frac{\chi}{24}-1)\Delta)
     \end{aligned}
 \end{equation*}
 i.e.,
  \begin{equation*}
  \mathcal{F}^{(1,1)}=\frac{1}{2}S^{jk}\Delta_{jk}-\mathcal{F}^{(1,0)}_j\Delta^j+\frac{1}{2}C_{jkl}S^{kl}\Delta^l-(\frac{\chi}{24}-1)\Delta+hol.amb.
 \end{equation*}
 or
   \begin{equation}\label{eq:2.26}
  \mathcal{F}^{(1,1)}=-\mathcal{F}^{(1,0)}_z\Delta^z-(\frac{\chi}{24}-1)\Delta+hol.amb.
 \end{equation}
 for one-parameter models
 
 In the two-loop level, the non-orientable diagram contribution satisifies the holomorphic anomaly equation,
 \begin{equation*}
 \begin{aligned}
     \partial_{\bar{i}}\mathcal{K}^{(1,1)}&=\frac{1}{2}C^{Pjk}_{\bar{i}} \Delta_{jk}-\mathcal{K}_j\Delta^j_{\bar{i}}\\
     &=\partial_{\bar{i}}(\frac{1}{2}S^{Pjk}\Delta_{jk}-\mathcal{K}_j\Delta^j)+\frac{1}{2}S^{Pjk}C_{jkl}\Delta^{l}_{\bar{i}}+(\frac{1}{2}C_{jkl}C^{Pkl}_{\bar{i}}-G_{\bar{i}j})\Delta^j\\
    & =\partial_{\bar{i}}(\frac{1}{2}S^{Pjk}\Delta_{jk}-\mathcal{K}_j\Delta^j+\frac{1}{2}C_{jkl}^{Pkl}\Delta^j-\Delta)
 \end{aligned}
 \end{equation*}
 i.e.,
  \begin{equation*}
   \mathcal{K}^{(1,1)} =\frac{1}{2}S^{Pjk}\Delta_{jk}-\mathcal{K}_j\Delta^j+\frac{1}{2}C_{jkl}^{Pkl}\Delta^j-\Delta+hol.amb.
 \end{equation*}
 and,
 \begin{equation}\label{eq:2.27}
     \mathcal{K}^{(1,1)}=-\mathcal{K}_z\Delta^z-\Delta+hol.amb.
 \end{equation}
 for one-parameter models. 
 
Then, the two-loop open/unoriented partition function has the expansion,
\begin{equation}
    i(\mathcal{F}^{(0,3)}+\mathcal{F}^{(1,1)}+\mathcal{K}^{(1,1)})=2\sum_{\substack{\text{d odd}\\ \text{k odd}}}(n_2^{(2,real)}-\frac{1}{24}n_d^{(0,real)})d^{kd/2},
\end{equation}
counting real BPS invariants.

 \section{Amplitudes and BPS Invariants on CICY Threefolds}
 In this section , we study open topological string amplitudes and BPS invariants on CICY threefolds in Weighted Projective Space. The Calabi-Yau  threefolds can be realized by Batyrev-Borisov toric mirror construction\cite{Batyrev1996,Batyrev1994,Borisov1993}. Relevant geometric data are listed in the following table \ref{tab:1} \cite{Almkvist2005,Batyrev1995}.
 \begin{table}[H]
\centering
 \begin{tabular}{l|lcccc}
 $X$&$\omega_0$&$W(0)$&$\mu$&$C_3$&$C_2 \cdot H$\\
 \hline
 $X_{4,4}\subset \mathbb{P}(1,1,1,1,2,2)$&$\sum_{n=0}^\infty\frac{(4n!)^2}{(n!)^4(2n!)^2}z^n$&$4$&$2^{12}$&$-144$&$40$\\[10pt]
 $X_{6,6}\subset \mathbb{P}(1,1,2,2,3,3)$&$\sum_{n=0}^\infty\frac{(6n!)^2}{(n!)^2(2n!)^2(3n!)^2}z^n$&$1$&$2^83^6$&$-120$&$22$\\[10pt]
 $X_{3,4}\subset \mathbb{P}(1,1,1,1,1,2)$&$\sum_{n=0}^\infty\frac{(3n!)(4n!)}{(n!)^5(2n!)^2}z^n$&$6$&$2^63^3$&$-156$&$48$\\[10pt]
 $X_{4,6}\subset \mathbb{P}(1,1,1,2,2,3)$&$\sum_{n=0}^\infty\frac{(6n!)(4n!)}{(n!)^3(2n!)^2(3n!)}z^n$&$2$&$2^{10}3^3$&$-156$&$32$\\
 \end{tabular}
 \caption{Geomereic Data of Complete Intersection in Weighted Projective Space }
 \label{tab:1}
  \vspace*{1ex}
 \begin{minipage}{\textwidth}
	$\varpi_0$ are the fundamental periods of the manifolds $X$. $W(0)$ and $\mu$ are related to the Yukawa couplings \ref{eq:2.18}. $C_3$ and $C_2\cdot H$ are the characteristic classes.
    \end{minipage}
 \end{table}

 \subsection{\texorpdfstring{$X_{4,4}$}{X(4,4)} }
The A-model manifold $X^*_{4,4}$ is given by the complete intersection of two quartics in the weighted projective space $\mathbb{P}^5_{(1,1,2,1,1,2)}$. $X^*_{4,4}$ is described  by a five dimensional polyhedron $\Delta^*$ with vertices,
\begin{equation*}
\begin{gathered}
   \nu_1^*=(-1,-2,-1,-1,-2),\quad\nu_2^*=(1,0,0,0,0),\quad \nu_3^*=(0,1,0,0,0),\\
    \nu_4^*=(0,0,1,0,0), \quad \nu_5^*=(0,0,0,1,0),\quad \nu_6^*=(0,0,0,0,1),
\end{gathered}
\end{equation*}
 and a inner point $\nu_0^*=(0,0,0,0,0)$. The nef-partition, $E_1=\{\nu_1^*,\nu_2^*,\nu_3^*\},E_2=\{\nu_4^*,\nu_5^*,\nu_6^*\}$, leads  a linear relation between vertices,
 \begin{equation*}
     l=(-4,-4;1,1,2,1,1,2).
\end{equation*}
and the equations of the mirror Calabi-Yau $X_{4,4}$,
 \begin{equation*}
 \begin{aligned}
     P_1&=a_{1,0}-a_1(X_2X_3^2X_4X_5X_6^2)^{-1}-a_2X_2-a_3X_3,\\
      P_2&=a_{2,0}-a_4X_4-a_5X_5-a_6X_6,
\end{aligned}
 \end{equation*}
 or
 \begin{equation*}
 \begin{aligned}
     P_1&=x_1^4+x_2^4+x_3^2+\psi x_4x_5x_6\\
      P_2&=x_4^4+x_5^4+x_6^2+\psi x_1x_2x_3,
\end{aligned}
 \end{equation*}
 after coordinate transform. 
 
 The period integral satisfies the Picard-Fuchs equation,
 \begin{equation*}
     \mathcal{L}_{PF}=\theta^4-2^4 z(4\theta+1)^2(4\theta+3)^2,
 \end{equation*}
 where $z=\psi^{-4}$ is the coordinate of $M_{CS}(X_{4,4})$, $\theta=z\frac{d}{dz}$ is the logarithmic derivative. The logarithmic solution underlies the mirror map from the complex structure moduli space of $X_{4,4}$,$M_{CS}(X_{4,4})$, to the Kahler moduli space of $X^*_{4,4}$, $M_{K}(X^*_{4,4})$
 \begin{equation*}
     z(q)=q - 960 q^2 + 213600 q^3 - 160471040 q^4 - 136981068240 q^5+\ldots.
 \end{equation*}
 
The curve $C$ wrapped by D-brane is defined as the intersection of $X_{4,4}$ with two planes,
 \begin{equation*}
    h_1=\{x_1+\alpha x_2=0\}, \quad h_2=\{x_4+\beta x_5=0\},\quad \alpha^4=\beta^4=-1
 \end{equation*}
 which splits into 
 \begin{equation*}
 \begin{aligned}
     C_+&=P_1\cap P_2 \cap \{x_3=0,x_6=0\},\\[10pt]
     C_-&=P_1\cap P_2 \cap \{x_3^2-\psi \beta x_4^6 x_6=0,x_6^2-\psi \alpha x_2^2x_3=0, x_3^3+\psi^2\beta^2\alpha x_2^2x_4^4=0\},
    \end{aligned}
\end{equation*}

The tree level has been solve in \cite{Walcher2009}. The domainwall tension is given by shifting $n\rightarrow n+\frac{1}{2}$ in the fundamental period,
\begin{equation*}
    \mathcal{T}(z)=\varpi(z;1/2),
\end{equation*}
with $\varpi(z;\rho)$ the generating function of the solutions of GKZ-system, and satisfies the inhomogeneous Picard-Fuchs equation,
\begin{equation}
    \mathcal{L}_{PF}\mathcal{T}=\frac{4}{(2\pi i)^2}z^{1/2},
\end{equation}

Given the domainwall tension,
\begin{equation*}
 \begin{aligned}
  &\mathcal{T}(q)= 64 q^{1/2} + \frac{50176}{9} q^{3/2} + \frac{116721664}{25} q^{5/2} + \frac{
 275837288448}{49} q^{7/2} \\
 &+\frac{671623092863488}{81} q^{9/2} +\ldots,\\
 \end{aligned}
 \end{equation*}
 the disk two point function $\Delta_{zz}$ is obtained by  equation \ref{eq:2.20}
\begin{equation}
    -i \Delta_{zz}=16 q^{1/2} + 12544 q^{3/2} + 29180416 q^{5/2} + 
   68959322112 q^{7/2}+\ldots. 
\end{equation}
Another crucial physical quantity is the Yukawa coupling given by equation \ref{eq:2.18},
\begin{equation}
     C_{zzz}=4 + 3712 q + 7863424 q^2 + 18453913600 q^3 + \ldots.
\end{equation}
Inserting $\Delta_{zz}$ and $C_{zzz}$ into \ref{eq:2.22}, the amplitudes $\mathcal{F}^{(0,2)}_z$ with one insertion is obtained,
\begin{equation}\label{eq:3.3}
     \mathcal{F}^{(0,2)}_{z}=-32 q - 20480 q^2 - 54477824 q^3 - 128899612672 q^4 + \ldots.
\end{equation}
and the partition function with zero genus and two boundaries is solved by direct integration,
\begin{equation}
     \mathcal{F}^{(0,2)}=-32 q - 10240 q^2 - \frac{54477824}{3}  - 32224903168 q^4  +\ldots
\end{equation}

In one-loop level, the Klein bottle partition function $\mathcal{K}$ contribution has to be considered to cancel the tadpole. In the holomorphic limit, $\mathcal{K}$ is given by equation\ref{eq:2.23}, with discriminant $1-2^{12} z$ from table \ref{tab:1}.
\begin{equation}
    \mathcal{K}= 32 q + 79456 q^2 + \frac{597450752}{3} q^3 + 467690079328 q^4  + \ldots.
\end{equation}

The sum of $\mathcal{F}^{(0,2)}$ and $\mathcal{K}$ encodes the genus one BPS invariants by equation \ref{eq:2.24}. The first a few invariants are listed in the following table \ref{tab:n144}.

The next loop level, $\mathcal{F}^{(0,3)}$,$\mathcal{F}^{(1,1)}$, and $\mathcal{K}^{(1,1)}$ need to be considered.
To begin with, $\mathcal{F}^{(0,3)}$ is solved by equation \ref{eq:2.25}. 
\begin{equation*}
\begin{aligned}
   & \mathcal{F}^{(0,3)}=-\frac{128}{3} q^{3/2} - \frac{63488}{3} q^{5/2} - \frac{205033472}{3} q^{7/2}-\frac{481914650624}{3} q^{9/2}\\ &- \frac{1180059100638208}{3} q^{11/2}-979023701838217216  q^{13/2}\\&-\frac{7394674719238803193856}{3}q^{15/2}\ldots.
\end{aligned}
\end{equation*}
Here $\mathcal{F}^{(0,2)}_z$ is from equation \ref{eq:3.3} and $\Delta^z$ is from \ref{eq:2.15}.

Secondly, $\mathcal{F}^{(1,1)}$ can be obtained from equation \ref{eq:2.26},
\begin{equation*}
\begin{aligned}
  &\mathcal{F}^{(1,1)}=-\frac{62}{3} q^{1/2} + 13792 q^{3/2} + 35328 q^{5/2} - \frac{4535729152}{3} q^{7/2}-9352462105744  q^{9/2} \\&- 31075040152947456  q^{11/2} -
 \frac{273946358860176045056}{3}  q^{13/2}\\&-255818331892779093696512q^{15/2}+\ldots,
\end{aligned}
\end{equation*}
Here we use the formula of $\mathcal{F}^{(1,0)}$ under the holomorphic limit\ref{eq:2.21}. The discriminant, Euler characteristic, and second Chern class can be read from table\ref{tab:1}.

Moreover, the unoriended contribution $\mathcal{F}^{(1,1)}$ is given by solving  equation \ref{eq:2.27},
\begin{equation*}
\begin{aligned}
 & \mathcal{K}^{(1,1)}= 2 q^{1/2} - 736 q^{3/2} + 532736 q^{5/2} + 2120819712 q^{7/2}  +6893532191312  q^{9/2}\\& + 20533504042332928  q^{11/2} + 
 58911806918364210176  q^{13/2}\\&+165605905671703558823936  q^{15/2}+\ldots.
\end{aligned}
\end{equation*}

The domainwall tension has another form When the D-brane wraps different curve. It leads to different partition function that are summarized in Appendix \ref{app:b} .

 \subsection{\texorpdfstring{$X_{6,6}$}{X(6,6)} }
 The description of A-model geometry $X_{6,6}^*$ in weighted projective space $ \mathbb{P}(1,2,3,1,2,3)$ is given by the polyhedron  $\Delta^*$ with one inner point  $\nu_0^*=(0,0,0,0,0)$ and the following six vertices,
\begin{equation*}
\begin{gathered}
   \nu_1^*=(-2,-3,-1,-2,-3),\quad \nu_2^*=(1,0,0,0,0),\quad \nu_3^*=(0,1,0,0,0),\\ \nu_4^*=(0,0,1,0,0), \quad \nu_5^*=(0,0,0,1,0),\quad \nu_6^*=(0,0,0,0,1),
\end{gathered}
\end{equation*}
satisfying a linear relation $l$ corresponding to the maximal trangulation of $\Delta^*$.
 \begin{equation*}
     l=(-6,-6;1,2,3,1,2,3).
\end{equation*}
By the nef-partition  $E_1=\{\nu_1^*,\nu_2^*,\nu_3^*\},E_2=\{\nu_4^*,\nu_5^*,\nu6^*\}$, the equations of mirror  threefold $X_{6,6}$ are obtained,
 \begin{equation*}
 \begin{aligned}
     P_1&=a_{1,0}-a_1(X_2^2X_3^3X_4X_5^2X_6^3)^{-1}-a_2X_2-a_3X_3,\\
      P_2&=a_{2,0}-a_4X_4-a_5X_5-a_6X_6,
\end{aligned}
 \end{equation*}
 and after homogenization, the equations are,
 \begin{equation*}
 \begin{aligned}
     P_1&=x_1^6+x_2^3+x_3^2+\psi x_4x_5x_6\\
      P_2&=x_4^6+x_5^3+x_6^2+\psi x_1x_2x_3.
\end{aligned}
 \end{equation*}
.
 
The Picard-Fuchs equation determines the deformations of the complex structure of $X_{6,6}$, 
 \begin{equation*}
     \mathcal{L}_{PF}=\theta^4-2^4 3^2z(6\theta+1)^2(6\theta+5)^2,
 \end{equation*}
with $z=\psi^{-6}$, which solves the mirror map,
\begin{equation*}
   z= q - 37440 q^2 + 84900960 q^3 - 15150231951360 q^4 +\ldots.
\end{equation*}
 
On $X_{6,6}$,  B-brane wraps on the curve $C$  defined by the intersection between $X_{6,6}$ and $h_1,h_2$,
 \begin{equation*}
    h_1=\{x_1^3+\alpha x_3=0\}, \quad h_2=\{x_4^3+\beta x_6=0\},\quad \alpha^2=\beta^2=-1.
 \end{equation*}
 
The   domainwall tension is given  in \cite{Walcher2009},
\begin{equation*}
\begin{aligned}
   &\mathcal{T}(q)= 256 q^{1/2} + \frac{15683584}{9} q^{3/2} + \frac{1626659168256}{25} q^{5/2} + \frac{
 186494133791883264}{49} q^{7/2}\\& + \frac{
 22204094064940507334656}{81} q^{9/2}+\ldots,
 \end{aligned}
\end{equation*} 
which satisfies the inhomogeneous Picard-Fuchs equation,
\begin{equation}\label{eq:3.7}
\mathcal{L}_{PF}\mathcal{T}=\frac{16}{(2\pi i)^2}z^{1/2}
\end{equation}

The two point function $\Delta_{zz}$ is given by  equation \ref{eq:2.20},
\begin{equation}
    -i \Delta_{zz}= 64 q^{1/2} + 3920896 q^{3/2} + 406664792064 q^{5/2}  +\ldots
\end{equation}
and the Yukawa coupling is given by equation \ref{eq:2.18},
\begin{equation}
  C_{zzz}=1 + 67104 q + 6778372896 q^2 + 771747702257664 q^3
  + \ldots.
\end{equation}

In the one-loop level, we need to consider $\mathcal{F}^{(0,2)}$ and $\mathcal{K}$.
The partition function $\mathcal{F}^{(0,2)}$ is solved by inserting  $\Delta_{zz}$ and $C_{zzz}$ into \ref{eq:2.22},
\begin{equation}
     \mathcal{F}^{(0,2)}=-2048 q - 56754176 q^2 - \frac{12214287269888}{3} q^3 + \ldots.
\end{equation}

The Klein bottle contribution $\mathcal{K}$ can be obtained  by equation\ref{eq:2.23},with discriminant $1-2^8 3^6z$.
\begin{equation}
     \mathcal{K}= 4608 q + 336967776 q^2 + 33802444071936 q^3  + \ldots.
\end{equation}

 The genus one BPS invariants on $X_{6,6}$ are extracted by equation  in table \ref{tab:n166}.

For the two-loop level, $\mathcal{F}^{(0,3)}$ is given by equation \ref{eq:2.25}. 
\begin{equation*}
\begin{aligned}
   & \mathcal{F}^{(0,3)}=-\frac{131072}{3} q^{3/2} - 2166358016 q^{5/2} - 
 245016521867264 q^{7/2}\\
 &- \frac{85168343643311833088}{3} q^{9/2} - 
 3396141356895699624525824 q^{11/2}\\&-413870034048732435400420753408  q^{13/2} +\ldots.
\end{aligned}
\end{equation*}
Here $\mathcal{F}^{(0,2)}_z$ is from equation \ref{eq:3.3} and $\Delta^z$ is from \ref{eq:2.15}.

Moreover, $\mathcal{F}^{(1,1)}$ and $\mathcal{K}^{(1,1)}$ can be obtained by solving the corresponding holomorphic anomaly equation \ref{eq:2.26}\ref{eq:2.27},
\begin{equation*}
\begin{aligned}
  &\mathcal{F}^{(1,1)}=-\frac{752}{3} q^{1/2} + \frac{12262144}{3} q^{3/2} + 44974083072 q^{5/2} + 
 3415441373093888 q^{7/2}\\
 &+ \frac{960915772990614614656}{3} q^{9/2} + 
 36672842106979528199706624 q^{11/2}\\& +4611989289188843900932102619136  q^{13/2} \ldots,\\[10pt]
 & \mathcal{K}^{(1,1)}= 32 q^{1/2} - 265728 q^{3/2} + 36233883648 q^{5/2} + 
 5892952067162112 q^{7/2}\\
 &+ 856257228798248625408 q^{9/2} + 
 119070042311934462996393984 q^{11/2} \\&+ 16222224902976953255052967624704  q^{13/2}+
  \ldots.
  \end{aligned}
\end{equation*}

Another two form of domainwall tension on $X_{6,6}$ are studied in \cite{Walcher2009}. We summarize the new amplitudes related to the new domainwall tension and omit certain detail in Appendix \ref{app:c}.

\subsection{\texorpdfstring{$X_{3,4}$}{X(3,4)} }
The A-incarnation  $X^*_{3,4} \subset\mathbb{P}^5_{(1,1,1,1,1,2)}$ is related to the  polyhedron $\Delta^*$ in the mirror construction.  There is one inner point $\nu^*_0=(0,0,0,0,0)$, and six vertices,
\begin{equation*}
\begin{gathered}
    \nu_1^*=(-1,-1,-1,-1,-2),\quad\nu_2^*=(1,0,0,0,0),\quad \nu_3^*=(0,1,0,0,0),\\ \nu_4^*=(0,0,1,0,0),\quad \nu_5^*=(0,0,0,1,0),\quad \nu_6^*=(0,0,0,0,1),
\end{gathered}
\end{equation*}
and the  maximal triangulation of $\Delta^*$ corresponds to the charge vector,
\begin{equation}\label{eq:3.10}
     l=(-3,-4;1,1,1,1,1,2).
\end{equation}

The mirror threefold $X_{3,4}$ is defined by the equation,
 \begin{equation*}
 \begin{aligned}
     P_1&=a_{1,0}-a_1(X_2X_3X_4X_5X_6^2)^{-1}-a_2X_2-a_3X_3,\\
      P_2&=a_{2,0}-a_4X_4-a_5X_5-a_6X_6,
\end{aligned}
 \end{equation*}
 and the period integrals are annihilated by  the GKZ-operator obtained from $l$ \ref{eq:3.10},
 \begin{equation*}
     \mathcal{L}=\prod^{5}_{i=1}\frac{\partial}{\partial a_i}\frac{\partial}{\partial a_6}^2-\left(\frac{\partial}{\partial a_{1,0}}\right)^3\left(\frac{\partial}{\partial a_{2,0}}\right)^4,
 \end{equation*}
 In terms of the logarithmic derivetives,$\mathcal{L}$ can be rewritten as,
 \begin{equation*}
     \mathcal{L}=\theta^5 (2\theta)(2\theta-1)-z\prod_{i=1}^4\prod^3_{j=1}(4\theta+i)(3\theta+j),
 \end{equation*}
with $z=\frac{a_1a_2a_3a_4a_5a_6^2}{a_{1,0}^3a_{2,0}^4}$  the coordinate on $M_{CS}(X_{3,4})$. It can be reduced to the Picard-Fuchs equation,
 \begin{equation*}
     \mathcal{L}_{PF}=\theta^4-2^{2} 3 z(4\theta+1)(3\theta+1)(3\theta+2)(4\theta+3)
 \end{equation*}
 and solves the mirror map,
 \begin{equation*}
     z=q-420q^2+47070q^3-12722000q^4-3647205075q^5+\ldots.
 \end{equation*}

At tree level,  the domainwall tension satisfies the inhomogeneous Picard-Fuchs equation \cite{Alim2010} as before,
\begin{equation}
    \mathcal{L}_{PF}\mathcal{T}=-\frac{3}{8\pi^2}z^{1/2},
\end{equation}
and  $\mathcal{T}(q)$ is obtained by restricting the domainwall tension on $X_{12}\subset \mathbb{P}^4_{(1,2,3,3,3)}$ to the point $t_2=t_3=0$,
 \begin{equation*}
     \mathcal{T}(q)=48 q^{1/2} + \frac{4192}{3} q^{3/2} + \frac{13300848}{25} q^{5/2} + \frac{
 13037063136}{49} q^{7/2} +\ldots,
 \end{equation*}
 after inserting the mirror map.

The disk amplitude $\Delta_{zz}$ is related to the domainwall tension by  equation \ref{eq:2.20}, 
\begin{equation}
    -i \Delta_{zz}= 12 q^{1/2} + 3144 q^{3/2} + 3325212 q^{5/2}  +\ldots. 
\end{equation}
and the Yukawa coupling is given by equation \ref{eq:2.18},
\begin{equation}
    \begin{aligned}
     C_{zzz}=6 + 1944 q + 1790424 q^2 + 1748375280 q^3 + \ldots.
    \end{aligned}
\end{equation}
Given $\Delta_{zz}$ and $C_{zzz}$, $\mathcal{F}^{(0,2)}$ can be solved by \ref{eq:2.22},
\begin{equation}
     \mathcal{F}^{(0,2)}=-12 q - 1200 q^2 - 1038568 q^3 - 759633600 q^4   + \ldots.
\end{equation}
In addition, there are Klein bottle contribution in the one loop level,
\begin{equation}
     \mathcal{K}=6 q + 10674 q^2 + 12298308 q^3 + 12377069538 q^4 + \ldots.
\end{equation}
Then, the genus one BPS invariants are extracted from $\mathcal{F}^{(0,2)}+\mathcal{K}$ in table \ref{tab:n134}.

At two loop level, the partition functions $\mathcal{F}^{(0,3)}$,$\mathcal{F}^{(1,1)}$, and $\mathcal{K}^{(1,1)}$  can be computed as before by solving the corresponding holomorphic anomaly equations.

\begin{equation*}
\begin{aligned}
   & \mathcal{F}^{(0,3)}=-8 q^{3/2} - 1104 q^{5/2} - 1968216 q^{7/2} - 1887797648 q^{9/2}\\ &- 
 1917394366728 q^{11/2} - 1975452738939456 q^{13/2}\\&-2058121403895827040  q^{15/2}+\ldots,\\[10pt]
  &\mathcal{F}^{(1,1)}=-\frac{23}{2} q^{1/2} + 1967 q^{3/2} + \frac{210573}{2} q^{5/2} + 
 78910897 q^{7/2} + 46406184491 q^{9/2} \\&+ 34970231999763 q^{11/2} + \frac{
 62290344299535317}{2} q^{13/2}\\& +31248235175950182696  q^{15/2} + \ldots,\\[10pt]
&  \mathcal{K}^{(1,1)}= q^{1/2} - 174 q^{3/2} + 35877 q^{5/2} + 64858286 q^{7/2} + 
 90920039958 q^{9/2}\\ &+ 114240749617194 q^{11/2}+ 
 137210633017558493 q^{13/2}\\&+160866795930971392080  q^{15/2} +\ldots.
\end{aligned}
\end{equation*}

\subsection{\texorpdfstring{$X_{4,6}$}{X(4,6)} }
The  A-model geometry $X^*_{4,6}$ is related to the  polyhedron $\Delta^*$ with vertices,
\begin{equation*}
\begin{gathered}
   \nu_1^*=(-1,-1,-2,-2,-3),\quad \nu_2^*=(1,0,0,0,0),\quad \nu_3^*=(0,1,0,0,0),\\ \nu_4^*=(0,0,1,0,0),\quad\nu_5^*=(0,0,0,1,0),\quad \nu_6^*=(0,0,0,0,1),
\end{gathered}
\end{equation*}
in the  Batyrev-Borisov construction, which give rise to the mirror geometry described by 
\begin{equation*}
    \begin{aligned}
   & a_{1,0}-a_1(X_2X_3X_4^2X_5^2X_6^3)^{-1}+a_2X_2+a_4X_4=0,\\
    &a_{2,0}-a_3X_3-a_5X_5-a_6X_6=0.
    \end{aligned}
\end{equation*}
The charge vector,
 \begin{equation*}
     l=(-4,-6;1,1,1,2,2,3),
\end{equation*}
defines the GKZ operator $\mathcal{L}$ that can be reduced to Picard-Fuchs operator,
 \begin{equation*}
     \mathcal{L}_{PF}=\theta^4-2^{4} 3 z(6\theta+1)(4\theta+1)(4\theta+3)(6\theta+5),
 \end{equation*}
with $z=\frac{a_1a_2a_3a_4^2a_5^2a_6^3}{a_{1,0}^4a_{2,0}^6}$. The  mirror map is given by the logarithmic solution of $\mathcal{L}_{PF}$,
 \begin{equation*}
     z=q-6144q^2+ 6866784q^3 -48364795904q^4-347475565045200q^5+\ldots.
 \end{equation*}

 The domainwall tension on $X_{4,6}$ is obtained by restricting the domainwall tension on $X_{12}\subset \mathbb{P}^4_{(1,2,2,3,4)}$ at the singular locus in $M_{CS}(X_{12})$. It is   the solution of the inhomogeneous Picard-Fuchs equation\cite{Alim2010},
 \begin{equation}
     \mathcal{L}_{PF}\mathcal{T}=\frac{4}{(2\pi i)^2} z^{1/2},
 \end{equation}
 and  expressed in the $q$-coordinate as,
\begin{equation*}
  \mathcal{T}(q)= 128 q^{1/2} + \frac{874496}{9} q^{3/2} + \frac{13288624128}{25} q^{5/2}  + \ldots,
\end{equation*}

The Yukawa coupling $C_{zzz}$ and the disk two point function $\Delta_{zz}$,
\begin{equation}
\begin{aligned}
 C_{zzz}=& 2 + 15552 q + 223248960 q^2 + 3614882992128 q^3  + \ldots.\\[10pt]
    -i \Delta_{zz}=& 32 q^{1/2} + 218624 q^{3/2} + 3322156032 q^{5/2} +\ldots. 
  \end{aligned}
\end{equation}
solves the equation \ref{eq:2.22} and obtains the partition function $\mathcal{F}^{(0,2)}$,
\begin{equation}
    \begin{aligned}
     \mathcal{F}^{(0,2)}=&-256 q - 753664 q^2 - \frac{24806760448}{3} q^3 + \ldots.
    \end{aligned}
\end{equation}

 To extract BPS invariants, the Klein bottle partition function $\mathcal{K}$ contribution from  equation \ref{eq:2.23} has to be considered.
\begin{equation}
     \mathcal{K}=384 q + 5097312 q^2 + 79609445376 q^3   +\ldots.
\end{equation}
The first a few  genus one BPS invariants are listed in table \ref{tab:n146}.

For the two loop level, $\mathcal{F}^{(0,3)}$,$\mathcal{F}^{(1,1)}$, and $\mathcal{K}^{(1,1)}$ are solved by the corresponding holomorphic anomaly equation,
\begin{equation*}
\begin{aligned}
    &\mathcal{F}^{(0,3)}=-\frac{4096}{3} q^{3/2} - 6750208 q^{5/2} - 124076818432 q^{7/2}\\ &- 
 \frac{6104855294771200}{3} q^{9/2} - 34531769181309009920 q^{11/2}\\&-596379664973249358856192  q^{13/2} +\ldots\\[10pt]
  &\mathcal{F}^{(1,1)}=-\frac{244}{3} q^{1/2} + \frac{522176}{3} q^{3/2} - 318506752 q^{5/2} - 
\frac{ 21046447364096}{3} q^{7/2} \\&- \frac{434275724425389088}{3} q^{9/2} - 
 2749664281128057287168 q^{11/2}\\&-\frac{152981476421267688845656064}{3}q^{13/2}+\ldots,\\[10pt]
 & \mathcal{K}^{(1,1)}=8q^{1/2} - 16512 q^{3/2} + 138651648 q^{5/2} + 
 3429512359936 q^{7/2}\\ &+ 73806308934007104 q^{9/2} + 
 1489674404926065349632 q^{11/2} \\&+29218644683428988803919872  q^{13/2}+\ldots.
\end{aligned}
\end{equation*}

 \section{Summary and Conclusion}
 In this article, we study the open topological string  partition function of low genus and boundaries by solving the extended holomiorphic anomaly equation. For four one-modulus complete intersection Calabi-Yau threefolds, we compute the partition function, and BPS invariants by including the orientifold plane contribution.  In the calculation, we assume that the holomorphic ambiguities are zero in most cases. It seems that the assumption of holomorphic ambiguity as zero does not  give rise to the integer BPS invariants for the two loop level. In the furture, we will use the localization technique to verify our results, identifies the holomorphic ambiguities, and try to compute BPS invariants of higher genus. Also, it is worthwhile to study the open string amplitudes and BPS invariants on more complicated geometric background, like two-parameter Calabi-Yau hypersurfaces in toric varieties and complete intersection in weighted projective spaces. 
 
Furthermore, we hope to extend the approach of remodeling B-model for local Calabi-Yau threefolds \cite{Bouchard2008} to the compact Calabi-Yau geometry. In the local case, loading, this method is originally  obtained as a solution to the loop equations of matrix models, giving an explicit form for its open and closed amplitudes in terms of residue calculus on the spectral curve of the matrix model unambiguously.  We will try to generalize the formalism of mirror  curve in the B-model geometry to higher dimensional manfolds so that it applies directly  to compact Calabi-Yau threefolds. It may provide some hints or implication on finding holomorphic ambiguity for solving the holomorphic anomaly equation.
 
 \appendix

\section{Genus One BPS Invariants}
 \begin{table}[H]
\centering
 \begin{tabular}{c|l}
 $d$&$n_d^{(1,real)}$\\
 \hline
$2$&$0$\\
$4$&$34608$\\
$6$&$90495488$\\
$8$&$217732588080$\\
$10$&$519139865625600$\\
$12$&$1244429369521121008$\\
$14$&$3006720894671076040704$\\
$16$&$7321444729803039221389872$\\
$18$&$17953446932877098422752258560$\\
$20$&$44297245727128905301804000758672$\\
$22$&$109886598389085980224169122468220928$\\
$24$&$273879048043960416419367130685174742512$\\
$26$&$685441373723245253419305721507760898859008$\\
$28$&$1721746947633131166800387889507150878112640176$\\
$30$&$4338887285554338333593607708507980232083840516608$\\
$32$&$10965963996574241145002521057990564470746280768723504$\\
$34$&$27787284139820409740273784776125024930464211651719647232$\\
$36$&$70577433742673178934242350825488542706466337629896022356096$\\
 \end{tabular}
 \caption{Real BPS Invariants $n_d^{(1,real)}$ on $X_{4,4}^*$}
 \label{tab:n144}
 \end{table}
 
 \begin{table}[H]
\centering
\rotatebox{90}{
 \begin{tabular}{c|l}
 $d$&$n_d^{(1,real)}$\\
 \hline
$2$&$1280$\\
$4$&$140106800$\\
$6$&$14865507490560$\\
$8$&$1618366267878984240$\\
$10$&$180879246399441648493312$\\
$12$&$20637943177316354437275713520$\\
$14$&$2392752424285671726708540669594880$\\
$16$&$280938904554424832218336437482840382000$\\
$18$&$33322503987626461636383333029163640299256320$\\
$20$&$3985484966875635872252891080039572436585586076304$\\
$22$&$480000354728744360465611971501493264886844125644120320$\\
$24$&$58150620987427492471708770141308835484350552668720148891120$\\
$26$&$7080336576485783736588492713534135020651896830908916304714499840$\\
$28$&$865859554481709720298276080110159012438129283882801758005389628254640$\\
$30$&$10629116492978029650234309164206402552577728937717548331282618388216
6785280$\\
$32$&$1309203828363866866581081598706633623471010905846012860231575526195847\
1721441840$
 \end{tabular}}
 \caption{Real BPS Invariants $n_d^{(1,real)}$ on $X_{6,6}^*$}
 \label{tab:n166}
 \end{table}
 
 \begin{table}[H]
\centering
 \begin{tabular}{c|l}
 $d$&$n_d^{(1,real)}$\\
 \hline
 $2$&$-3$\\
$4$&$4737$\\
$6$&$5629871$\\
$8$&$5808717969$\\
$10$&$5833299344640$\\
$12$&$5849047597419579$\\
$14$&$5891041195079975079$\\
$16$&$5967814657032068338641$\\
$18$&$6080507782501626356757812$\\
$20$&$6228358922063613246678825600$\\
$22$&$6410400570932796039941119094847$\\
$24$&$6625980254770789737512663516966507$\\
$26$&$6874895369173424708708048755071562560$\\
$28$&$7157401533148708877822459482246852554579$\\
$30$&$7474184389248964662717428155183980031936826$\\
$32$&$7826326324564347587478671218568566261580459217$\\
$34$&$8215278529651988449131114534693789042732127972935$\\
$36$&$8642841129411377770003000151742474654966182322188388$\\
 \end{tabular}
 \caption{Real BPS Invariants $n_d^{(1,real)}$ on $X_{3,4}^*$}
 \label{tab:n134}
 \end{table}
 
 \begin{table}[H]
\centering
\rotatebox{90}{
 \begin{tabular}{c|l}
 $d$&$n_d^{(1,real)}$\\
 \hline
$2$&$64$\\
$4$&$2171824$\\
$6$&$35670262592$\\
$8$&$572119847810608$\\
$10$&$9264343811094049984$\\
$12$&$152008070011375287627120$\\
$14$&$2524239861846741083787784768$\\
$16$&$42347361041482559192621064600112$\\
$18$&$716550646088663055133099960033534976$\\
$20$&$12212640014878093727055039426586080874768$\\
$22$&$209432794162057905066976004068843293806233408$\\
$24$&$3610555956393932419362664034621338976108612559344$\\
$26$&$62530394810056915742075848828164170928495828847300288$\\
$28$&$1087289471705430471456914140078475721731390116433863314736$\\
$30$&$18972692030691869017569026158060107915758257894036779123617216$\\
$32$&$332100638118912607108732075831745677990230635273178077491923999280$\\
$34$&$58293751439773786276298974513519564386164982737392093498738081703850\
88$\\
$36$&$10257992459591667334465084196511068095716339806899158048539291997504\
9948288$\\
 \end{tabular}}
 \caption{Real BPS Invariants $n_d^{(1,real)}$ on $X_{4,6}^*$}
 \label{tab:n146}
 \end{table}
 
 \section{\texorpdfstring{Partition Functions on $X_{4,4}$}{X(4,4)} } \label{app:b}
With  the D-brane wrapping the following curve,
 \begin{equation*}
     X_{4,4}=\{P_1=0,P_2=0\} \cap h_1=\{x_1^2+\alpha_1\sqrt{2}x_3=0\}\cap h_2=\{x_4^2+\alpha_2\sqrt{2}x_6=0\},\alpha_1^2=\alpha_2^2=-1,
 \end{equation*}
 there is another  domainwall $\mathcal{T}(z)$ satisfying\cite{Walcher2009},
  \begin{equation*}
     \mathcal{L}_{PF}\mathcal{T}(z)=\frac{1}{(2\pi i)^2}(\frac{8}{27}\eta z^{1/3}+\frac{800}{27}\eta^2z^{2/3})
 \end{equation*}
 with $\eta^3=1$.
 and $\mathcal{T}$ is
 \begin{equation*}
 \begin{aligned}
 \mathcal{T}(z(q))&=\frac{2}{9}(\tilde{\eta}\varpi(z;1/3)+\tilde{\eta}^2\varpi(z,2/3))\\
     &=24 q^{1/3} + 150 q^{2/3} + \frac{2571}{2} q^{4/3} + \frac{417024}{25} q^{5/3} +\ldots,
\end{aligned}
 \end{equation*}
 with $\tilde{\eta}^3=1$.

 Several amplitudes are listed as follow,
   \begin{equation*}
     \begin{aligned}
     & F^{0,1}_{zz}=3 q^{1/3} + 75 q^{2/3} + 2571 q^{4/3} + 52128 q^{5/3} + 
 5677584 q^{7/3} + 131074059 q^{8/3} \\&+ 13380832800 q^{10/3} + 
 310078975968 q^{11/3} + 32533714689024 q^{13/3} \\&+ 
 756271662397200 q^{14/3}+ 80669533303699467 q^{16/3} + 
 1878508762455982080 q^{17/3}\\& + 202613133782293403616 q^{19/3} +\ldots,\\[10pt]
 &F^{0,2}=-\frac{27}{16} q^{2/3} - \frac{225}{4} q- \frac{16875}{32} q^{4/3} - \frac{ 10611}{20} q^{5/3} - \frac{140409}{8} q^2 - \frac{974700}{7} q^{7/3}\\& - \frac{ 49254507 }{64}q^{8/3} - \frac{126814971}{4} q^3 - \frac{
 13362659451}{40} q^{10/3} - \frac{14555738232 }{11}q^{11/3} \\&- \frac{
 899642497977}{16} q^4 - \frac{7818600973032 }{13}q^{13/3}- \frac{
 17720520118920}{7} q^{14/3} \\&- 109860797109960 q^5+\ldots.
     \end{aligned}
 \end{equation*}
 
 \begin{equation*}
     \begin{aligned}
& F^{0,3}=-\frac{9}{32} q - \frac{675}{32} q^{4/3} - \frac{16875}{32} q^{5/3} - \frac{
 36765}{8} q^2 - \frac{186651}{16} q^{7/3} - \frac{6599475}{32} q^{8/3} - \frac{
 48068955}{32} q^3 \\&- \frac{553202811 }{16}q^{10/3} - \frac{
 13023069651}{16} q^{11/3} - \frac{61206596433}{8} q^4 - \frac{
 1305111161691}{16} q^{13/3} \\&- \frac{30239942059107}{16} q^{14/3} - \frac{
 34877455807311}{2} q^5+\ldots,\\[10pt]
 &F^{1,1}=-3 q^{1/3} - \frac{475}{4} q^{2/3} + \frac{7431}{4} q^{4/3} + 90040 q^{5/3} + 
 78480 q^{7/3} - \frac{19724437}{4} q^{8/3} \\&- 398429040 q^{10/3} - 
 3346434248 q^{11/3} - 1944661693968 q^{13/3} -
 38405520795668 q^{14/3} +\ldots
     \end{aligned}
 \end{equation*}

 \section{\texorpdfstring{Partition Functions on $X_{6,6}$}{X(6,6)} }
  \label{app:c}
  \subsection{Domainwall Tension II}
  When the D-brane wraps the curve,
 \begin{equation*}
     X_{6,6}=\{P_1=0,P_2=0\} \cap h_1=\{x_1^2+2^{1/3}\alpha_1 x_2=0\}\cap h_2=\{x_4^2+\alpha_2 2^{1/3}x_5=0\},\alpha_1^3=\alpha_2^3=-1,
 \end{equation*}
 the domianwall tension satisifies the inhomogeneous  Picard-Fuchs equation\cite{Walcher2009},
 \begin{equation*}
     \mathcal{L}_{PF}\mathcal{T}(z)=\frac{1}{(2\pi i)^2}(\frac{2}{3}\tilde{\eta}z^{1/3}+216\tilde{\eta}^2 z^{2/3})
 \end{equation*}
 where $\tilde{\eta}$ dependens on  a combination of $\alpha_1$ and $\alpha_2$.
 Then, after inserting the mirror map, $\mathcal{T}(q)(\tilde{\eta}=1)$ is,
 \begin{equation*}
 \begin{aligned}
 \mathcal{T}(z(q))&=\frac{2}{9}(\tilde{\eta}\varpi(z;1/3)+\tilde{\eta}^2\varpi(z,2/3))\\
    &= 54 q^{1/3} + \frac{2187}{2} q^{2/3} + \frac{1733643}{8} q^{4/3} + \frac{
 252362304}{25} q^{5/3} +\ldots,
 \end{aligned}
 \end{equation*}
 with $\tilde{\eta}^3=1$
 
 The relevant amplitudes are listed as follow.
 
 \begin{equation*}
     \begin{aligned}
     & F^{0,1}_{zz}=6 q^{1/3} + 486 q^{2/3} + 385254 q^{4/3} + 28040256 q^{5/3} + 
 39380410656 q^{7/3} + 2970341239014 q^{8/3} \\&+ 
 4499651224412736 q^{10/3} + 341829582883713216 q^{11/3} + 
 535032321621406746624 q^{13/3}\\& + 40760047932529671669024 q^{14/3} + 
 64986912884772303566448870 q^{16/3} \\&+ 
 4958366424970308770413707264 q^{17/3} + 
 7999190717324822926707846353088 q^{19/3}  +\ldots,\\[10pt]
 &F^{0,2}=-27 q^{2/3} - 2916 q - \frac{177147 }{2}q^{4/3} - \frac{3310956}{5} q^{5/3} - 
 79899858 q^2 - \frac{17108148672}{7} q^{7/3}\\& - \frac{171633915387}{4} q^{8/3} - 
 5758204096908 q^3 - \frac{980289702999702 }{5}q^{10/3} - \frac{39356189371608192}{11} q^{11/3} \\&- 498396671394586857 q^4 - \frac{
 227301205949499102336}{13} q^{13/3}- \frac{
 2342879992102562008704}{7} q^{14/3}\\&- \frac{238001239476332766988416 }{5}q^5+\ldots,
     \end{aligned}
 \end{equation*}
  \begin{equation*}
     \begin{aligned}
& F^{0,3}=-36 q - 8748 q^{4/3} - 708588 q^{5/3} - 21234960 q^2 - 
 454073688 q^{7/3} - 32164935084 q^{8/3} \\&- 965557032876 q^3 - 
 50254848648792 q^{10/3} - 3807431288304408 q^{11/3}\\& - 
 122006194369498512 q^4 - 5813866460445646680 q^{13/3} - 
 441980260922360950488 q^{14/3}\\& - 14242379576529165875136 q^5+\ldots,\\[10pt]
 &F^{1,1}=-\frac{35}{2} q^{1/3} - \frac{4779}{2} q^{2/3} + \frac{381189}{2} q^{4/3} + 
 52806816 q^{5/3} + 3085343928 q^{7/3}\\& + \frac{741480466149}{2} q^{8/3} + 
 227008394824256 q^{10/3} + 34734669284988288 q^{11/3} \\&+ 
 23806991100654596112 q^{13/3} + 3021677505059070600792 q^{14/3} +\ldots
     \end{aligned}
 \end{equation*}

  \subsection{Domainwall Tension III}
  The equation \ref{eq:3.7} has another solution,
  \begin{equation*}
  \begin{aligned}
  \mathcal{T}(z(q))&=\frac{1}{4}(\tilde{\eta}\varpi(z;1/4)+\tilde{\eta}^2\varpi(z,1/2)+\tilde{\eta}^3 \varpi(z;3/4))\\
    & = 32 q^{1/4} + 256 q^{1/2} + \frac{25088}{9} q^{3/4} + \frac{
 2092032}{25} q^{5/4} +\ldots,
  \end{aligned}
  \end{equation*}
  with $\tilde{\eta}^4=1$, which is independent to the domainwall tension in Section 3.2.
  
  The relevant amplitudes are listed as follow.
 \begin{equation*}
     \begin{aligned}
     & F^{0,1}_{zz}=2 q^{1/4} + 64 q^{1/2} + 1568 q^{3/4} + 130752 q^{5/4} + 
 3920896 q^{3/2} + 87386112 q^{7/4} \\&+ 13297603088 q^{9/4} + 
 406664792064 q^{5/2} + 9382033536768 q^{11/4} + 
 1517266245622272 q^{13/4} \\&+ 46623533447970816 q^{7/2} + 
 1081732875754733568 q^{15/4} + 180314551186106647608 q^{17/4}\\& + 
 5551023516235126833664 q^{9/2} + 129090812443883028381312 q^{19/4}  +\ldots,\\[10pt]
 &F^{0,2}=-4 q^{1/2} - \frac{512}{3} q^{3/4} - 5184 q - \frac{401408}{5} q^{5/4} - \frac{
 2713216}{3} q^{3/2} - \frac{30482432}{7} q^{7/4} \\&- 141430784 q^2 - \frac{
 20026621952}{9} q^{9/4} - 27029641408 q^{5/2} - \frac{
 3192144662528}{11} q^{11/4} \\&- 10211216431104 q^3 - \frac{
 2258170042318848}{13} q^{13/4} - \frac{16057665232558080}{7} q^{7/2}\\&- \frac{
 366689651197018112}{15} q^{15/4} - 883989881776373760 q^4 - \frac{
 261571893681164648448}{17} q^{17/4}\\& - \frac{
 1872789489451577109856}{9} q^{9/2} - \frac{
 43691386492446963795968 }{19}q^{19/4}+\ldots,\\[10pt]
& F^{0,3}=-\frac{4}{3} q^{3/4} - 128 q - 7232 q^{5/4} - \frac{733184}{3} q^{3/2} - 
 5752448 q^{7/4} - 86075392 q^2 \\&- \frac{3079180288}{3} q^{9/4} - 
 11832852480 q^{5/2} - 254801522976 q^{11/4} - 3813556254720 q^3 \\&- 
 63257958588928 q^{13/4} - 1354065828249600 q^{7/2} - 
 30856030594393088 q^{15/4}\\& - 485588407814225920 q^4 - 
 7934104744264269824 q^{17/4} - \frac{470970231779873521664}{3} q^{9/2} \\&- 
 3586148378350321933936 q^{19/4}+\ldots,\\[10pt]
 &F^{1,1}=-\frac{29}{6} q^{1/4} - \frac{752 }{3}q^{1/2} - \frac{25480}{3} q^{3/4} + 
 40992 q^{5/4} + \frac{12262144}{3} q^{3/2} + 212925696 q^{7/4}\\& + \frac{
 2451671612}{3} q^{9/4} + 44974083072 q^{5/2} + 
 1110117057984 q^{11/4} + 63019510378496 q^{13/4}\\& + 
 3415441373093888 q^{7/2} + 127380480523266048 q^{15/4} + 
 7104502724515200414 q^{17/4}\\& + \frac{960915772990614614656}{3} q^{9/2}+\ldots
     \end{aligned}
 \end{equation*}

\printbibliography

@article{Griffiths1969a,
 author = {Phillip A. Griffiths},
 journal = {Annals of Mathematics},
 number = {3},
 pages = {496--541},
 publisher = {Annals of Mathematics},
 title = {On the Periods of Certain Rational Integrals: II},
 volume = {90},
 year = {1969}
}

@article{Griffiths1969,
 author = {Philip A. Griffiths},
 journal = {Annals of Mathematics},
 number = {3},
 pages = {460--495},
 publisher = {Annals of Mathematics},
 title = {On the Periods of Certain Rational Integrals: I},
 volume = {90},
 year = {1969}
}

@article{Griffiths1983,
     author = {Griffiths, Phillip A.},
     title = {Infinitesimal variations of hodge structure {(III)} : determinantal varieties and the infinitesimal invariant of normal functions},
     journal = {Compositio Mathematica},
     pages = {267--324},
     publisher = {Martinus Nijhoff Publishers},
     volume = {50},
     number = {2-3},
     year = {1983},
     zbl = {0576.14009},
     mrnumber = {720290}
   
}

@article{Green1989,
author = {Mark L. Green},
title = {{Griffiths' infinitesimal invariant and the Abel-Jacobi map}},
volume = {29},
journal = {Journal of Differential Geometry},
number = {3},
publisher = {Lehigh University},
pages = {545 -- 555},
year = {1989}
}

@article{Witten1997,
	year = 1997,
	month = {12},
	publisher = {Elsevier {BV}},
	volume = {507},
	number = {3},
	pages = {658--690},
	author = {Edward Witten},
	title = {Braves and the dynamics of {QCD}},
	journal = {Nuclear Physics B}
}

@article{Bershadsky1993a,
    author = "Bershadsky, M. and Cecotti, S. and Ooguri, H. and Vafa, C.",
    editor = "Greene, B. and Yau, Shing-Tung",
    title = "{Holomorphic anomalies in topological field theories}",
    eprint = "hep-th/9302103",
    archivePrefix = "arXiv",
    reportNumber = "HUTP-93-A008, RIMS-915",
    journal = "Nucl. Phys. B",
    volume = "405",
    pages = "279--304",
    year = "1993"
}

@article{Bershadsky1993,
    author = "Bershadsky, M. and Cecotti, S. and Ooguri, H. and Vafa, C.",
    title = "{Kodaira-Spencer theory of gravity and exact results for quantum string amplitudes}",
    eprint = "hep-th/9309140",
    archivePrefix = "arXiv",
    reportNumber = "HUTP-93-A025, RIMS-946, SISSA-142-93-EP",
    journal = "Commun. Math. Phys.",
    volume = "165",
    pages = "311--428",
    year = "1994"
}

@article{Yamaguchi2004,
    author = "Yamaguchi, Satoshi and Yau, Shing-Tung",
    title = "{Topological string partition functions as polynomials}",
    eprint = "hep-th/0406078",
    archivePrefix = "arXiv",
    journal = "JHEP",
    volume = "07",
    pages = "047",
    year = "2004"
}

@article{Huang2006,
    author = "Huang, Min-xin and Klemm, Albrecht and Quackenbush, Seth",
    title = "{Topological string theory on compact Calabi-Yau: Modularity and boundary conditions}",
    eprint = "hep-th/0612125",
    archivePrefix = "arXiv",
    reportNumber = "MAD-TH-06-12",
    journal = "Lect. Notes Phys.",
    volume = "757",
    pages = "45--102",
    year = "2009"
}

@article{Hosono2007,
    author = "Hosono, Shinobu and Konishi, Yukiko",
    title = "{Higher genus Gromov-Witten invariants of the Grassmannian, and the Pfaffian Calabi-Yau threefolds}",
    eprint = "0704.2928",
    archivePrefix = "arXiv",
    primaryClass = "math.AG",
    journal = "Adv. Theor. Math. Phys.",
    volume = "13",
    number = "2",
    pages = "463--495",
    year = "2009"
}

@article{Walcher2007,
    author = "Walcher, Johannes",
    title = "{Extended holomorphic anomaly and loop amplitudes in open topological string}",
    eprint = "0705.4098",
    archivePrefix = "arXiv",
    primaryClass = "hep-th",
    journal = "Nucl. Phys. B",
    volume = "817",
    pages = "167--207",
    year = "2009"
}

@article{Cook2007,
    author = "Cook, Paul L. H. and Ooguri, Hirosi and Yang, Jie",
    title = "{Comments on the Holomorphic Anomaly in Open Topological String Theory}",
    eprint = "0706.0511",
    archivePrefix = "arXiv",
    primaryClass = "hep-th",
    reportNumber = "CALT-68-2651",
    journal = "Phys. Lett. B",
    volume = "653",
    pages = "335--337",
    year = "2007"
}

@article{Konishi2007,
    author = "Konishi, Yukiko and Minabe, Satoshi",
    title = "{On solutions to Walcher's extended holomorphic anomaly equation}",
    eprint = "0708.2898",
    archivePrefix = "arXiv",
    primaryClass = "math.AG",
    journal = "Commun. Num. Theor. Phys.",
    volume = "1",
    pages = "579--603",
    year = "2007"
}

@article{Alim2007,
    author = "Alim, Murad and Lange, Jean Dominique",
    title = "{Polynomial Structure of the (Open) Topological String Partition Function}",
    eprint = "0708.2886",
    archivePrefix = "arXiv",
    primaryClass = "hep-th",
    reportNumber = "LMU-ASC-57-07",
    journal = "JHEP",
    volume = "10",
    pages = "045",
    year = "2007"
}

@article{Batyrev1995,
   year = 1995,
	month = {4},
  
	publisher = {Springer Science and Business Media {LLC}
},
  
	volume = {168},
  
	number = {3},
  
	pages = {493--533},
  
	author = {Victor V. Batyrev and Duco van Straten},
  
	title = {Generalized hypergeometric functions and rational curves on Calabi-Yau complete intersections in toric varieties},
  
	journal = {Communications in Mathematical Physics}
}

@article{Walcher2006,

	year = 2007,
	
	month = {10},
  
	publisher = {Springer Science and Business Media {LLC}
},
  
	volume = {276},
  
	number = {3},
  
	pages = {671--689},
  
	author = {Johannes Walcher},
  
	title = {Opening Mirror Symmetry on the Quintic},
  
	journal = {Communications in Mathematical Physics}
}

@article{Dijkgraaf2002,
    author = "Dijkgraaf, Robbert and Vafa, Cumrun",
    title = "{Matrix models, topological strings, and supersymmetric gauge theories}",
    eprint = "hep-th/0206255",
    archivePrefix = "arXiv",
    reportNumber = "HUTP-02-A028, ITFA-2002-22",
    journal = "Nucl. Phys. B",
    volume = "644",
    pages = "3--20",
    year = "2002"
}

@article{Dijkgraaf2002a,
    author = "Dijkgraaf, Robbert and Vafa, Cumrun",
    title = "{On geometry and matrix models}",
    eprint = "hep-th/0207106",
    archivePrefix = "arXiv",
    reportNumber = "HUTP-02-A030, ITFA-2002-24",
    doi = "10.1016/S0550-3213(02)00764-2",
    journal = "Nucl. Phys. B",
    volume = "644",
    pages = "21--39",
    year = "2002"
}

@article{Dijkgraaf2002b,
    author = "Dijkgraaf, Robbert and Vafa, Cumrun",
    title = "{A Perturbative window into nonperturbative physics}",
    eprint = "hep-th/0208048",
    archivePrefix = "arXiv",
    reportNumber = "HUTP-02-A034, ITFA-2002-34",
    month = "8",
    year = "2002"
}

@article{Marino2008,

 	year = 2008,
	month = {1},
  
	publisher = {Springer Science and Business Media {LLC}
},
  
	volume = {2008},
  
	number = {03},
  
	pages = {060--060},
  
	author = {Marcos Mari{\~{n}}o},
  
	title = {Open string amplitudes and large order behavior in topological string theory},
  
	journal = {Journal of High Energy Physics}
}

@article{Eynard2007,
    author = "Eynard, Bertrand and Orantin, Nicolas",
    title = "{Invariants of algebraic curves and topological expansion}",
    eprint = "math-ph/0702045",
    archivePrefix = "arXiv",
    reportNumber = "SPHT-07-021",
    journal = "Commun. Num. Theor. Phys.",
    volume = "1",
    pages = "347--452",
    year = "2007"
}

@inproceedings{Marino2004,
    author = "Marino, Marcos",
    title = "{Les Houches lectures on matrix models and topological strings}",
    eprint = "hep-th/0410165",
    archivePrefix = "arXiv",
    reportNumber = "CERN-PH-TH-2004-199",
    month = "10",
    year = "2004"
}

@article{Bouchard2008,
	year = 2008,
	month = {9},
	publisher = {Springer Science and Business Media {LLC}
},
  
	volume = {287},
  
	number = {1},
  
	pages = {117--178},
  
	author = {Vincent Bouchard and Albrecht Klemm and Marcos Mari{\~{n}}o and Sara Pasquetti},
  
	title = {Remodeling the B-Model},
  
	journal = {Communications in Mathematical Physics}
}

@article{Eynard2007a,
    author = "Eynard, Bertrand and Marino, Marcos and Orantin, Nicolas",
    title = "{Holomorphic anomaly and matrix models}",
    eprint = "hep-th/0702110",
    archivePrefix = "arXiv",
    reportNumber = "CERN-PH-TH-2007-031, SPHT-07-020",
    journal = "JHEP",
    volume = "06",
    pages = "058",
    year = "2007"
}

@article{Eynard2014,
    author = "Eynard, B",
    title = "{A short overview of the ''Topological recursion''}",
    eprint = "1412.3286",
    archivePrefix = "arXiv",
    primaryClass = "math-ph",
    reportNumber = "IPHT-T14-033-CRM3335",
    month = "12",
    year = "2014"
}

@article{Pandharipande2008,

  
  
	year = 2008,
	month = {2},
  
	publisher = {American Mathematical Society ({AMS})},
  
	volume = {21},
  
	number = {4},
  
	pages = {1169--1209},
  
	author = {R. Pandharipande and J. Solomon and J. Walcher},
  
	title = {Disk enumeration on the quintic 3-fold},
  
	journal = {Journal of the American Mathematical Society}
	}

@article{Morrison2007,
    author = "Morrison, David R. and Walcher, Johannes",
    title = "{D-branes and Normal Functions}",
    eprint = "0709.4028",
    archivePrefix = "arXiv",
    primaryClass = "hep-th",
    journal = "Adv. Theor. Math. Phys.",
    volume = "13",
    number = "2",
    pages = "553--598",
    year = "2009"
}

@article{Krefl2008,
    author = "Krefl, Daniel and Walcher, Johannes",
    title = "{Real Mirror Symmetry for One-parameter Hypersurfaces}",
    eprint = "0805.0792",
    archivePrefix = "arXiv",
    primaryClass = "hep-th",
    reportNumber = "CERN-PH-TH-2008-091, LMU-ASC-24-08",
    journal = "JHEP",
    volume = "09",
    pages = "031",
    year = "2008"
}

@article{Walcher2009,
    author = "Walcher, Johannes",
    title = "{Calculations for Mirror Symmetry with D-branes}",
    eprint = "0904.4905",
    archivePrefix = "arXiv",
    primaryClass = "hep-th",
    reportNumber = "CERN-PH-TH-2009-056",
    journal = "JHEP",
    volume = "09",
    pages = "129",
    year = "2009"
}

@article{Knapp2008,
    author = "Knapp, Johanna and Scheidegger, Emanuel",
    title = "{Towards Open String Mirror Symmetry for One-Parameter Calabi-Yau Hypersurfaces}",
    eprint = "0805.1013",
    archivePrefix = "arXiv",
    primaryClass = "hep-th",
    reportNumber = "MPP-2008-42",
    journal = "Adv. Theor. Math. Phys.",
    volume = "13",
    number = "4",
    pages = "991--1075",
    year = "2009"
}

@article{Walcher2007a,
    author = "Walcher, Johannes",
    title = "{Evidence for Tadpole Cancellation in the Topological String}",
    eprint = "0712.2775",
    archivePrefix = "arXiv",
    primaryClass = "hep-th",
    journal = "Commun. Num. Theor. Phys.",
    volume = "3",
    pages = "111--172",
    year = "2009"
}

@article{Shimizu2011,
	year = 2011,
	month = {3},
  
	publisher = {Springer Science and Business Media {LLC}
},
  
	volume = {2011},
  
	number = {3},
  
	author = {Masahide Shimizu and Hisao Suzuki},
  
	title = {Open mirror symmetry for pfaffian Calabi-Yau 3-folds},
  
	journal = {Journal of High Energy Physics}
}

@article{Krefl2009,
    author = "Krefl, Daniel and Walcher, Johannes",
    title = "{The Real Topological String on a local Calabi-Yau}",
    eprint = "0902.0616",
    archivePrefix = "arXiv",
    primaryClass = "hep-th",
    reportNumber = "CERN-PH-TH-2009-022, LMU-ASC-06-09",
    month = "2",
    year = "2009"
}

@article{Bouchard2010,
	year = 2010,
	month = {3},
  
	publisher = {Springer Science and Business Media {LLC}
},
  
	volume = {296},
  
	number = {3},
  
	pages = {589--623},
  
	author = {Vincent Bouchard and Albrecht Klemm and Marcos Mari{\~{n}}o and Sara Pasquetti},
  
	title = {Topological Open Strings on Orbifolds},
  
	journal = {Communications in Mathematical Physics}
}

@article{Kontsevich1994,
    author = "Kontsevich, Maxim",
    title = "{Homological Algebra of Mirror Symmetry}",
    eprint = "alg-geom/9411018",
    archivePrefix = "arXiv",
    month = "11",
    year = "1994"
}

@inproceedings{Kontsevich2000,
    author = "Kontsevich, Maxim and Soibelman, Yan",
    title = "{Homological mirror symmetry and torus fibrations}",
    booktitle = "{KIAS Annual International Conference on Symplectic Geometry and Mirror Symmetry}",
    eprint = "math/0011041",
    archivePrefix = "arXiv",
    pages = "203--263",
    month = "11",
    year = "2000"
}

@article{Candelas1990,
    author = "Candelas, Philip and De La Ossa, Xenia C. and Green, Paul S. and Parkes, Linda",
    title = "{A Pair of Calabi-Yau manifolds as an exactly soluble superconformal theory}",
    reportNumber = "UTTG-25-90",
    journal = "Nucl. Phys. B",
    volume = "359",
    pages = "21--74",
    year = "1991"
}

@article{Eynard2012,
    author = "Eynard, Bertrand and Orantin, Nicolas",
    title = "{Computation of Open Gromov\textendash{}Witten Invariants for Toric Calabi\textendash{}Yau 3-Folds by Topological Recursion, a Proof of the BKMP Conjecture}",
    eprint = "1205.1103",
    archivePrefix = "arXiv",
    primaryClass = "math-ph",
    reportNumber = "IPHT-T12-030",
    journal = "Commun. Math. Phys.",
    volume = "337",
    number = "2",
    pages = "483--567",
    year = "2015"
}

@article{Katz2001,
    author = "Katz, Sheldon H. and Liu, Chiu-Chu Melissa",
    editor = "Auckly, David and Bryan, Jim",
    title = "{Enumerative geometry of stable maps with Lagrangian boundary conditions and multiple covers of the disc}",
    eprint = "math/0103074",
    archivePrefix = "arXiv",
    doi = "10.2140/gtm.2006.8.1",
    journal = "Adv. Theor. Math. Phys.",
    volume = "5",
    pages = "1--49",
    year = "2001"
}

@article{Li2001,
    author = "Li, Jun and Song, Yun S.",
    editor = "Auckly, David and Bryan, Jim",
    title = "{Open string instantons and relative stable morphisms}",
    eprint = "hep-th/0103100",
    archivePrefix = "arXiv",
    reportNumber = "SLAC-PUB-8736, SU-ITP-01-08",
    journal = "Adv. Theor. Math. Phys.",
    volume = "5",
    pages = "67--91",
    year = "2001"
}

@article{Kachru2000,

  
	year = 2000,
	month = {11},
  
	publisher = {American Physical Society ({APS})},
  
	volume = {62},
  
	number = {12},
  
	author = {Shamit Kachru and Sheldon Katz and Albion Lawrence and John McGreevy},
  
	title = {Mirror symmetry for open strings},
  
	journal = {Physical Review D}
}

@article{Kachru1999,
  
	year = 2000,
	month = {6},
  
	publisher = {American Physical Society ({APS})},
  
	volume = {62},
  
	number = {2},
  
	author = {Shamit Kachru and Sheldon Katz and Albion Lawrence and John McGreevy},
  
	title = {Open string instantons and superpotentials},
  
	journal = {Physical Review D}
}

@inproceedings{Witten1993,
    author = "Witten, Edward",
    title = "{Quantum background independence in string theory}",
    booktitle = "{Conference on Highlights of Particle and Condensed Matter Physics (SALAMFEST)}",
    eprint = "hep-th/9306122",
    archivePrefix = "arXiv",
    reportNumber = "IASSNS-HEP-93-29",
    month = "6",
    year = "1993"
}

@article{Batyrev1996,
    author = "Batyrev, V. V. and Borisov, L. A.",
    editor = "Greene, B. and Yau, Shing-Tung",
    title = "{Dual cones and mirror symmetry for generalized Calabi-Yau manifolds}",
    journal = "AMS/IP Stud. Adv. Math.",
    volume = "1",
    pages = "71--86",
    year = "1996"
}

@ARTICLE{Borisov1993,
       author = {{Borisov}, Lev},
        title = "{Towards the Mirror Symmetry for Calabi-Yau Complete intersections in Gorenstein Toric Fano Varieties}",
      journal = {arXiv e-prints},
         year = 1993,
        month = oct,
       eprint = {alg-geom/9310001},
 primaryClass = {math.AG}
}

@article{Batyrev1994,
    author = "Batyrev, Victor V. and Borisov, Lev A.",
    title = "{On Calabi-Yau complete intersections in toric varieties}",
    eprint = "alg-geom/9412017",
    archivePrefix = "arXiv",
    month = "12",
    year = "1994"
}

@article{Cecotti1993,
	year = 1993,
	month = {10},
  
	publisher = {Springer Science and Business Media {LLC}
},
  
	volume = {157},
  
	number = {1},
  
	pages = {139--178},
  
	author = {Sergio Cecotti and Cumrun Vafa},
  
	title = {Ising model {andN}=2 supersymmetric theories},
  
	journal = {Communications in Mathematical Physics}
}

@article{Li2009,
    author = "Li, Si and Lian, Bong H. and Yau, Shing-Tung",
    title = "{Picard-Fuchs Equations for Relative Periods and Abel-Jacobi Map for Calabi-Yau Hypersurfaces}",
    eprint = "0910.4215",
    archivePrefix = "arXiv",
    primaryClass = "math.AG",
    month = "10",
    year = "2009"
}

@article{Aganagic2000,
    author = "Aganagic, Mina and Vafa, Cumrun",
    title = "{Mirror symmetry, D-branes and counting holomorphic discs}",
    eprint = "hep-th/0012041",
    archivePrefix = "arXiv",
    reportNumber = "HUTP-00-A047",
    month = "12",
    year = "2000"
}

@article{Aganagic2001,
    author = "Aganagic, Mina and Klemm, Albrecht and Vafa, Cumrun",
    title = "{Disk instantons, mirror symmetry and the duality web}",
    eprint = "hep-th/0105045",
    archivePrefix = "arXiv",
    reportNumber = "HUTP-01-A023, HU-EP-01-21",
    journal = "Z. Naturforsch. A",
    volume = "57",
    pages = "1--28",
    year = "2002"
}

@Inbook{Green1994,
author="Green, Mark L.",
editor="Bardelli, Fabio
and Albano, Alberto",
title="Infinitesimal Methods in Hodge Theory",
bookTitle="Algebraic Cycles and Hodge Theory: Lectures given at the 2nd Session of the Centro Internationale Matematico Estivo (C.I.M.E.) held in Torino, Italy, June 21-29, 1993",
year="1994",
publisher="Springer Berlin Heidelberg",
address="Berlin, Heidelberg",
pages="1--92",
isbn="978-3-540-49046-3"
}

@book{voisin2002,
place={Cambridge}, 
series={Cambridge Studies in Advanced Mathematics}, 
title={Hodge Theory and Complex Algebraic Geometry I},
volume={1}, 
publisher={Cambridge University Press}, 
author={Voisin, Claire}, 
editor={Schneps, LeilaTranslator}, 
year={2002},
collection={Cambridge Studies in Advanced Mathematics}
}

@book{voisin2003,
place={Cambridge}, 
series={Cambridge Studies in Advanced Mathematics},
title={Hodge Theory and Complex Algebraic Geometry II}, 
volume={2}, 
publisher={Cambridge University Press}, 
author={Voisin, Claire},
editor={Schneps, LeilaTranslator}, 
year={2003}, 
collection={Cambridge Studies in Advanced Mathematics}
}

@article{Morrison1993,
  title={Mirror symmetry and rational curves on quintic threefolds: a guide for mathematicians},
  author={Morrison, David R},
  journal={Journal of the American Mathematical Society},
  volume={6},
  number={1},
  pages={223--247},
  year={1993}
}

@article{Schmid1973,
author = {Schmid, Wilfried},
journal = {Inventiones mathematicae},
pages = {211-320},
title = {Variation of Hodge Structure: The Singularities of the Period Mapping.},
volume = {22},
year = {1973},
}

@article{Gelfand1990,
  title={Generalized Euler integrals and A-hypergeometric functions},
  author={Gel'fand Im and Mikhail M. Kapranov and Andrei Zelevinsky},
  journal={Advances in Mathematics},
  year={1990},
  volume={84},
  pages={255-271}
}

@article{Hosono1995,
    author = "Hosono, S. and Lian, B. H. and Yau, Shing-Tung",
    title = "{GKZ generalized hypergeometric systems in mirror symmetry of Calabi-Yau hypersurfaces}",
    eprint = "alg-geom/9511001",
    archivePrefix = "arXiv",
    journal = "Commun. Math. Phys.",
    volume = "182",
    pages = "535--578",
    year = "1996"
}

@article{Hosono1993,
    author = "Hosono, S. and Klemm, A. and Theisen, S. and Yau, Shing-Tung",
    title = "{Mirror symmetry, mirror map and applications to Calabi-Yau hypersurfaces}",
    eprint = "hep-th/9308122",
    archivePrefix = "arXiv",
    reportNumber = "HUTMP-93-0801, LMU-TPW-93-22",
    journal = "Commun. Math. Phys.",
    volume = "167",
    pages = "301--350",
    year = "1995"
}

@article{Hosono1994,
    author = "Hosono, S. and Klemm, A. and Theisen, S. and Yau, Shing-Tung",
    editor = "Greene, B. and Yau, Shing-Tung",
    title = "{Mirror symmetry, mirror map and applications to complete intersection Calabi-Yau spaces}",
    eprint = "hep-th/9406055",
    archivePrefix = "arXiv",
    reportNumber = "HUTMP-94-02, CERN-TH-7303-94, LMU-TPW-94-03",
    journal = "Nucl. Phys. B",
    volume = "433",
    pages = "501--554",
    year = "1995"
}

@article{Zamolodchikov1986,
    author = "Zamolodchikov, A. B.",
    title = "{Irreversibility of the Flux of the Renormalization Group in a 2D Field Theory}",
    journal = "JETP Lett.",
    volume = "43",
    pages = "730--732",
    year = "1986"
}

@article{Mayr2001,
    author = "Mayr, Peter",
    title = "{N=1 mirror symmetry and open / closed string duality}",
    eprint = "hep-th/0108229",
    archivePrefix = "arXiv",
    reportNumber = "CERN-TH-2001-230",
    doi = "10.4310/ATMP.2001.v5.n2.a1",
    journal = "Adv. Theor. Math. Phys.",
    volume = "5",
    pages = "213--242",
    year = "2002"
}

@article{Lerche2001,
    author = "Lerche, W. and Mayr, P.",
    title = "{On N=1 mirror symmetry for open type 2 strings}",
    eprint = "hep-th/0111113",
    archivePrefix = "arXiv",
    reportNumber = "CERN-TH-2001-301",
    month = "11",
    year = "2001"
}

@article{Lerche2002,
    author = "Lerche, W. and Mayr, P. and Warner, N.",
    title = "{Holomorphic N=1 special geometry of open - closed type II strings}",
    eprint = "hep-th/0207259",
    archivePrefix = "arXiv",
    reportNumber = "CERN-TH-2002-174",
    month = "7",
    year = "2002"
}

@article{Lerche2002a,
    author = "Lerche, W. and Mayr, P. and Warner, N.",
    title = "{N=1 special geometry, mixed Hodge variations and toric geometry}",
    eprint = "hep-th/0208039",
    archivePrefix = "arXiv",
    reportNumber = "CERN-TH-2002-175",
    month = "8",
    year = "2002"
}

@article{Dijkgraaf1990,
    author = "Dijkgraaf, Robbert and Verlinde, Herman L. and Verlinde, Erik P.",
    title = "{Topological strings in d \ensuremath{<} 1}",
    reportNumber = "PUPT-1204, IASSNS-HEP-90-71",
    journal = "Nucl. Phys. B",
    volume = "352",
    pages = "59--86",
    year = "1991"
}

@article{Forbes2003,
    author = "Forbes, Brian",
    title = "{Open string mirror maps from Picard-Fuchs equations on relative cohomology}",
    eprint = "hep-th/0307167",
    archivePrefix = "arXiv",
    month = "7",
    year = "2003"
}

@article{Forbes2005,
    author = "Forbes, Brian and Jinzenji, Masao",
    title = "{Extending the Picard-Fuchs system of local mirror symmetry}",
    eprint = "hep-th/0503098",
    archivePrefix = "arXiv",
    journal = "J. Math. Phys.",
    volume = "46",
    pages = "082302",
    year = "2005"
}

@article{Almkvist2005,
  title={Tables of Calabi--Yau equations},
  author={Almkvist, Gert and Van Enckevort, Christian and Van Straten, Duco and Zudilin, Wadim},
  journal={arXiv preprint math/0507430},
  year={2005}
}

@article{Alim2010,
    author = "Alim, Murad and Hecht, Michael and Jockers, Hans and Mayr, Peter and Mertens, Adrian and Soroush, Masoud",
    title = "{Type II/F-theory Superpotentials with Several Deformations and N=1 Mirror Symmetry}",
    eprint = "1010.0977",
    archivePrefix = "arXiv",
    primaryClass = "hep-th",
    reportNumber = "LMU-ASC-74-10, SU-ITP-10-28, NSF-KITP-10-117",
    journal = "JHEP",
    volume = "06",
    pages = "103",
    year = "2011"
}
\end{document}